\documentclass[fleqn,usenatbib,useAMS]{mnras}

\usepackage{graphicx}	
\usepackage{amsmath}	
\usepackage{amssymb}	
\usepackage{multicol}        
\usepackage{array}
\usepackage{multirow}
\usepackage{bm}		
\usepackage{pdflscape}	
\usepackage{color}
\usepackage{comment}
\usepackage{widetext}
\usepackage[version=4]{mhchem}
\usepackage[normalem]{ulem}
\usepackage{lineno}

\DeclareRobustCommand{\VAN}[3]{#2}
\let\VANthebibliography\thebibliography
\def\thebibliography{\DeclareRobustCommand{\VAN}[3]{##3}\VANthebibliography}

\newcommand{\Jf}{$\mathcal{J}$}
\newcommand{\Fermi}{{\it Fermi}}
\newcommand{\unit}[1]{{\;\mathrm{#1}}}

\definecolor{forestgreen}{rgb}{0.13, 0.545, 0.13}

\definecolor{applegreen}{rgb}{0.55, 0.71, 0.0}

\definecolor{orange}{rgb}{1, 0.65, 0.0}

\providecommand{\tabularnewline}{\\}

\usepackage[T1]{fontenc}
\usepackage{ae,aecompl}
\graphicspath{{Figures/}}

\usepackage{newtxtext,newtxmath}

\title[]{How do the dynamics of the Milky Way - Large Magellanic Cloud system affect gamma-ray constraints on particle dark matter?}

\author[C.~Eckner et al.]{Christopher Eckner$^{1}$\thanks{Contact e-mails: \href{mailto:}{eckner@lapth.cnrs.fr, calore@lapth.cnrs.fr, d.erkal@surrey.ac.uk, s.lilleengen@surrey.ac.uk, michael.petersen@roe.ac.uk}}, 
Francesca Calore$^{1}$, 
Denis Erkal$^{2}$,
Sophia~Lilleengen$^{2}$, 
Michael S. Petersen$^{3,4}$  
\\ 
$^{1}$LAPTh, CNRS, USMB, F-74940 Annecy, France\\
$^{2}$ Department of Physics, University of Surrey, Guildford GU2 7XH, UK\\
$^3$CNRS and Sorbonne Universite, UMR 7095, Institut d'Astrophysique de Paris, 98 bis Boulevard Arago, F-75014 Paris, France\\
$^4$Institute for Astronomy, University of Edinburgh, Royal Observatory, Blackford Hill, Edinburgh EH9 3HJ, UK\\ 
}

\pubyear{2022}

\begin{document}

\label{firstpage}
\pagerange{\pageref{firstpage}--\pageref{lastpage}}
\maketitle

\begin{abstract}
Previous studies on astrophysical dark matter (DM) constraints have all assumed that the Milky Way's (MW) DM halo can be modelled in isolation. However, recent work suggests that the MW’s largest dwarf satellite, the Large Magellanic Cloud (LMC), has a mass of 10-20$\%$ that of the MW and is currently merging with our Galaxy. As a result, the DM haloes of the MW and LMC are expected to be strongly deformed.
We here address and quantify the impact of the
dynamical response caused by the passage of the LMC through the MW on the prospects for indirect DM searches.
Utilising a set of state-of-the-art numerical simulations of the evolution of the MW-LMC system, we derive the DM distribution in both galaxies at the present time based on the Basis Function Expansion formalism. Consequently, we build \Jf-factor all-sky maps of the MW-LMC system to study the impact of the LMC passage on gamma-ray indirect searches for thermally produced DM annihilating in the outer MW halo as well as within the LMC halo standalone. We conduct a detailed analysis of 12 years of \Fermi-LAT data that incorporates various large-scale gamma-ray emission components and we quantify the systematic uncertainty associated with the imperfect knowledge of the astrophysical gamma-ray sources. We find that the dynamical response caused by the LMC passage can alter the constraints on the velocity-averaged annihilation cross section for weak scale particle DM at a level comparable to the existing observational uncertainty of the MW halo's density profile and total mass. 
\end{abstract}





\section{Introduction}
\label{sec:intro}

The Large Magellanic Cloud (LMC) is believed to be on its first approach to the Milky Way (MW) since the early Universe \citep{Besla+2007}. This is supported by many lines of evidence which show that it still hosts a massive dark matter (DM) halo, in line with expectations from abundance matching, $\sim 2\times10^{11}~{\rm M}_\odot$ \citep[e.g.][]{Behroozi+2013,Moster+2013}. First, the nearby presence of the Small Magellanic Cloud (SMC) requires an LMC mass of $\sim10^{11}~{\rm M}_\odot$ in order to remain bound to the LMC \citep{Kallivayalil+2013}. Similar analyses show that the recently discovered Magellanic satellites also require similarly high LMC masses, $\sim (1-2)\times 10^{11}~{\rm M}_\odot$ \citep{Erkal+2020,Patel+2020} in order to have originally been bound to the LMC. The LMC is massive enough that it induces a strong reflex motion in the MW \citep{Gomez+2015}, as evidenced by its effect on the timing argument with Andromeda and the nearby Hubble flow, which require a mass of $\sim 2.5\times10^{11}~{\rm M}_\odot$ \citep{Penarrubia+2016}. This reflex motion is also seen in the MW's stellar halo, both in its kinematics \citep{Erkal+2021,Petersen+2021} and density \citep{Belokurov+2019,Conroy+2021}, all requiring a mass $>\sim10^{11}~{\rm M}_\odot$. Finally, its effect has been seen and characterized in many stellar streams around the MW, giving masses of $(1.3-1.8)\times10^{11}~{\rm M}_\odot$ \citep{2019MNRAS.487.2685E,Koposov+2019,Shipp+2019,Shipp+2021,Vasiliev+2021}.

Such a massive LMC halo is $\sim(10-20)\%$ of the MW's mass \citep[e.g.][]{Wang+2020}, suggesting that this merger should have a substantial effect on the dark matter haloes of both galaxies. This has been studied with simulations \citep[e.g.][]{Laporte+2018,Garavito-Camargo+2019,Garavito-Camargo+2021,Petersen+2020} which have shown that substantial DM deformations are expected. 

Several works have also explored the observational consequences of these deformations. \cite{Vasiliev+2021} and \cite{2022arXiv220501688L} showed that these deformations can affect the Sagittarius and Orphan-Chenab stream, respectively. These effects are larger than current observational uncertainties, suggesting that in the future it will be possible to measure these deformations. 
\cite{Conroy+2021} claimed the detection of the MW-LMC haloes' deformations through a
positive correlation between giant halo stars in \textit{Gaia} and \textit{WISE} data and the simulations from~\cite{Garavito-Camargo+2019}.
Such deformations can also alter the velocity distribution of DM. 
\citet{2019JCAP...11..013B} and \citet{Donaldson+2022} studied the impact of the MW-LMC dynamics on the local DM velocity distribution, finding an enhanced reach of direct DM detection experiments.

In this work, for the first time, we will explore another avenue for characterising these deforming haloes more directly: gamma-ray searches for DM signals.
In the standard paradigm, 80\% of the matter content of the Universe,
i.e.~the DM, can be explained by new particles beyond the ones
in the Standard Model of particle physics.
In particular, weak scale DM particle candidates 
may be thermally produced in the early universe through interactions
with Standard Model particles.
These so-called Weakly Interacting Massive Particles (WIMPs) are
still present today, and can self-annihilate or decay 
into final stable products contributing 
to the fluxes of cosmic gamma rays and charged cosmic rays~(see, e.g., the review on this topic in \cite{Cirelli:2012tf}).
The spatial distribution of gamma rays from DM annihilation or decay in the sky
depends on the distribution of DM in the target of interest, 
through the integral along the line of sight (l.o.s) 
of the DM density (squared in the case of annihilation).
Therefore, any change in the 
expected DM spatial density, $\rho$, affects 
the large-scale {\it morphology} of the DM signal.
In this respect, a perfectly legitimate question to ask 
is: How does the MW-LMC dynamics affect gamma-ray searches for 
DM?
We quantify the answer to this question 
by analysing all-sky data of the Fermi Large Area Telescope (\Fermi-LAT) and
search for DM at high Galactic latitudes.
This approach follows traditionally
performed searches for DM with \Fermi-LAT data~\citep{2012ApJ...761...91A, Zechlin:2017uzo, Chang:2018bpt},
and builds on the modelling and optimisation
of the astrophysical background
and foreground components of the gamma-ray high-energy sky.

In addition, we also study the impact of our DM
spatial model on the constraints derived from gamma-ray
observations of a region centred around the LMC, 
to be compared with previous \Fermi-LAT analyses~\citep{Buckley:2015doa}, 
and the recent spatial model from~\cite{Regis:2021glv}.

The main novelties of the present work are:
(i) the modelling of the
DM gamma-ray all-sky signal based on state-of-the-art simulations
of the MW-LMC interaction~\citep{Donaldson+2022,2022arXiv220501688L}; 
and (ii) the quantification of the uncertainty on the
DM limits issued by one of the most accurate models of the MW potential
and its associated uncertainties~\citep{2017MNRAS.465...76M},
based on which we can properly assess 
 the impact of the MW-LMC  dynamics in gamma-ray DM searches.
As a result, our limits on DM at high latitude represent 
the most up-to-date and robust constraints from \Fermi-LAT
gamma-ray observations.

The paper is organised as follows:
In Sec.~\ref{sec:simulation}, we describe
the modelling of the DM signal, namely
its spatial distribution, based on the outcome of 
simulations of the MW-LMC interaction, and 
spectrum. 
We dedicate Sec.~\ref{sec:fermianalysis} to explaining the details of the \Fermi-LAT analysis, statistical framework and
fitting procedure for the high-latitude sky and LMC regions. 
This section is complemented by Appendix~\ref{app:astro_comp} 
about the modelling of the astrophysical fore- and back-ground
gamma-ray components. 
In Sec.~\ref{sec:upperlimits}, we present the new constraints on 
the particle DM parameter space from the high-latitude and LMC regions.
We discuss the impact of varying the interstellar emission model,
i.e.~the dominant source of background modelling systematic uncertainties in 
Appendix~\ref{app:iem_limits}.
We draw our conclusions in Sec.~\ref{sec:conclusions}.


\section{Milky Way -- Large Magellanic Cloud dark matter distribution}
\label{sec:simulation}

\subsection{Simulation of the Milky Way -- Large Magellanic Cloud dynamics}
\label{sec:mw-lmc-sim}

In order to explore the deformations of the MW and LMC, we use a suite of $N$-body simulations run with the \textsc{exp} code \citep{Petersen+2022}. Unlike other $N$-body codes which evaluate forces with a hierarchical tree \citep[e.g.][]{Appel1985,Barnes+1986} or a particle mesh \citep[e.g.][]{Klypin+1983,White+1983}, this code evolves $N$-body particles by using a basis function expansion (BFE). In particular, \textsc{exp} uses biorthogonal basis functions for the potential and density which satisfy the Poisson equation. The angular structure of each model is described by spherical harmonics, with harmonic indices $\ell$ and $|m|\le\ell$ ($\ell=0$ is the monopole, $\ell=1$ is the dipole, $\ell=2$ is the quadrupole, and so on), while the radial structure is described by $n$ basis functions per harmonic order derived from the Poisson equation \citep{Weinberg1999}. At each timestep, the coefficients of each basis function are estimated by summing each potential term in the expansion over the location of the particles \citep[see eq. 5 in][]{Petersen+2022}. With these coefficients, the forces can be readily computed by differentiating the potential. This technique has several advantages for our study (see \citealt{Petersen+2022} for more details). First, it is computationally efficient, scaling as $\mathcal{O}\propto N$ where $N$ is the number of particles used, allowing for more particles at a reduced computational cost. Second, the forces are less noisy than standard gravity solvers, allowing us to study the subtleties of the MW and LMC deformations without worrying about noise. Lastly, the BFE of the density allows us to quickly determine the density when modelling gamma-ray signals.

In this work, we make use of two sets of simulations of the ongoing MW-LMC merger. First, we use the simulation of \cite{2022arXiv220501688L}. Their MW and LMC system is based on the results of \cite{2019MNRAS.487.2685E} who measured the MW and LMC potentials with the Orphan stellar stream. In particular, the MW is initialized as a Miyamoto-Nagai disc \citep{Miyamoto+1975} with a mass of $6.8\times10^{10}~{\rm M}_\odot$, a scale radius of $3$ kpc, and a scale height of $0.28$ kpc, a Hernquist bulge \citep{hernquist90} with a mass of $5\times10^{9}~{\rm M}_\odot$ and a scale radius of $0.5$ kpc, and an Navarro-Frenk-White (NFW) halo \citep{NFW+1996} with a mass of $7.92\times10^{11}~{\rm M}_\odot$, a scale radius of $12.8$ kpc, and a concentration of 15.3. The LMC is modelled as a Hernquist DM halo with a mass of $1.25
\times10^{11}~{\rm M}_\odot$ and a scale radius of $14.9$ kpc. These values all come from the best-fit model of \cite{2019MNRAS.487.2685E} assuming a spherical DM halo for the MW. As a result, we dub this simulation the `Erkal19' model. \cite{2022arXiv220501688L} show that the MW and LMC DM haloes experience strong deformations, most notably in the dipole of the MW and in the quadrupole of the LMC.

Second, we use the simulation suite of \cite{Donaldson+2022} which uses the same \textsc{exp} technique and considers four different MW-LMC models. These simulations are built to roughly match the rotation curve and total mass constraints of the MW \citep{Eilers+2019,Eadie+2019} and LMC \citep{vanderMarel:2013jza,Penarrubia+2016,2019MNRAS.487.2685E}. The MW model consists of an exponential disc with a mass of $5\times10^{10}~{\rm M}_\odot$, a scale radius of 3 kpc, and a sech$^2$ scale height of 0.6 kpc, and a dark matter halo with a profile given by $\rho(r) = \rho_0(r)\tilde{r}^{-1}(1+\tilde{r})^{-\alpha}T(r)$ with scaled radius $\tilde{r}=r/R_s$ where $R_s$ is the scale radius, and the truncation function $T(r)=0.5\left(1-{\rm erf}\left[(r-r_{\rm trunc})/w_{\rm trunc}\right]\right)$ with $r_{\rm trunc}=430$ kpc and $w_{\rm trunc}=54$ kpc. One can set this profile to be either an NFW ($\alpha=2$) or Hernquist ($\alpha=3$) profile dark matter halo. We require a similar mass enclosed at 50 kpc by tuning the respective scale radii of the MW models: $R_{s,{\rm NFW MW}}=15$ kpc and $R_{s,{\rm Hernquist MW}}=44$ kpc. The total mass of the NFW halo is $1\times10^{12}~{\rm M}_\odot$, while the total mass of the Hernquist halo is $0.94\times10^{12}~{\rm M}_\odot$. We build two LMCs, modelled as a dark matter halo only, again following the same truncated halo profile as above, with $R_{s,{\rm NFW LMC}}=33.8$ kpc and $R_{s,{\rm Hernquist LMC}}=63$ kpc. The total masses are set to be $2.5\times10^{11}~{\rm M}_\odot$ for the NFW LMC and $2.35\times10^{11}~{\rm M}_\odot$ for the Hernquist LMC.

The simulation suite is constructed as a grid of four models by mixing the MW and LMC halo profiles. That is, one simulation is an NFW MW and NFW LMC, one is an NFW MW and Hernquist LMC, one is a Hernquist MW and NFW LMC, and one is a Hernquist MW and Hernquist LMC. We label this second set of simulations by referring to the halo profiles of the MW and the LMC at the stage of simulation initialisation, i.e.~either an NFW or a Hernquist (HERN) profile. 
When analyzing the gamma-ray signals expected from these models, we take an approach inspired by \cite{2022arXiv220501688L} and consider the full multipole expansions as well as the monopoles for comparison.
The monopole terms describe the spherical representation of the MW and LMC haloes, but changes over time as the coefficients of the monopole radial basis functions vary. Due to the relatively short travel times of gamma rays through the Milky Way ($\sim1$ Myr to travel 300 kpc) compared to the timescales over which the basis function changes substantially \citep[$\sim 50-100$ Myr, see fig. 1 of][]{2022arXiv220501688L}, we only consider the coefficients at the present-day.
By comparing these two, we can see how much the deformations affect the predicted gamma-ray signal. We consider this difference as the most robust result of this work.
Indeed, as a word of caution, we notice that,
while the initial models of \cite{2022arXiv220501688L} and \cite{Donaldson+2022} were consistent with the MW gravitational potential constraints, the full consistency with 
the MW potential has not been
{\it a posteriori} checked for the finally deformed models. That is, the rotation curve and mass enclosed constraints originally imposed may no longer be met.

Therefore, in order to quantify the impact of the absolute value of 
the new constraints for gamma-ray DM searches,
in addition to these two sets of models, we also consider a static MW model. For this potential, we use the results of \cite{2017MNRAS.465...76M} who modelled the MW with a bulge, four discs (thin, thick, H\textsc{i}, and H$_2$), and an NFW DM halo. \cite{2017MNRAS.465...76M} fit this model to a range of data and constraints: maser data, the solar velocity, terminal velocity curves, the vertical force near the plane of the disc, and a mass constraint based on satellite kinematics (see Sec.~3 of \citealt{2017MNRAS.465...76M} for more details). While these constraints are primarily within the plane of the MW disc, it represents one of the most accurate models of the MW potential. 
To explore how the uncertainties in the MW potential affect the predicted gamma-ray signal, we sample over the posterior chains from \cite{2017MNRAS.465...76M}. 
This allows us to (a) quantify the systematic uncertainties
on the high-latitude DM limits from the MW gravitational potential,
and (b) properly assess the impact of the variations of the limits
induced by the MW-LMC dynamics.

\subsection{Dark matter-induced gamma-ray signal}

In this work, we consider the gamma-ray emission resulting from pair-annihilating thermally produced DM in the MW and LMC halo. We restrict ourselves to the prompt gamma-ray component of these interactions neglecting potential secondary or tertiary contributions from particle cascades triggered by the primary annihilation products. The expected (prompt) differential gamma-ray flux $\textrm{d}\Phi_{\gamma}/\textrm{d}E_{\gamma}/\mathrm{d}\Omega$  at the top of the Earth's atmosphere reads (see, e.g., \citealt{Cirelli:2010xx, Bringmann:2012ez})
\begin{equation}
\label{DMflux}
  \frac{d\Phi_{\gamma}}{d\Omega\, dE_\gamma} (E_\gamma,\psi) = 
  \left(\vphantom{\frac{dN_\gamma^{f}}{dE_\gamma}\sum_f}\frac{1}{4\pi} \int_\mathrm{l.o.s}
  d\ell(\psi) \rho_\chi^2(\bm{r})\right) 
  \left({\frac{\langle\sigma v\rangle_\mathrm{ann}}{2S_\chi m_{\chi}^2} \sum_f
  B_f\frac{dN_\gamma^{f}}{dE_\gamma}}\right) \,,
\end{equation}
where $\psi$ refers to the direction of the line-of-sight in Galactic coordinates and $E_{\gamma}$ quantifies the gamma rays' energy. The DM-induced gamma-ray flux factors into two contributions under the assumption of velocity-independent annihilation cross section -- the so-called $s$-wave annihilation process. The term in the first parenthesis is commonly referred to as \Jf-factor while the term in the second parenthesis includes and describes the particle physics model chosen for the DM candidate under study. In what follows, we provide further details about the ingredients required to compute both contributions to the DM gamma-ray signal.

\subsubsection{\Jf-factor all-sky maps for the Milky Way}
\label{sec:jfactor_maps}

In order to produce \Jf-factor maps of the MW that incorporate deformations induced by the dynamics of the MW-LMC system, we take the density from the present-day snapshots of the simulations in \cite{2022arXiv220501688L} and \cite{Donaldson+2022}, and measure the square of the DM density $\rho_{\chi}$ along lines of sight on a HEALPix grid \citep{2005ApJ...622..759G} with $N_{\rm side}=64$, resulting in 49,152 lines of sight. Along these lines of sight, the density is evaluated in the centre of 1 kpc-sized bins out to 100 kpc. This choice of resolution is not a limitation of the BFE, which can faithfully reproduce structure on $\sim100$ pc scales, but rather motivated by the chosen resolution for the \Fermi-LAT gamma-ray data set (see Sec.~\ref{sec:lat_data_selection}). For these mock observations, the Sun is placed at a distance of $8.249$ kpc from the Galactic centre \citep{GravityCollaboration2020} in the plane of the MW disc. As discussed in Section~\ref{sec:mw-lmc-sim}, we create \Jf-factor maps for the full BFE as well as solely the monopole term for the MW standalone or the combined MW-LMC system. 
The outlined extraction scheme allows us to directly perform the l.o.s.~integral of $\rho_{\chi}^2$ where the l.o.s.~direction is given by the central coordinates of a particular HEALPix pixel. To this end, we linearly interpolate the density values at the extracted positions and numerically integrate the result from 0 to 100 kpc.

In contrast, we derive two-dimensional all-sky \Jf-factor maps of the static MW model of \cite{2017MNRAS.465...76M} with the publicly available software \textsc{CLUMPY} (version 3) \cite{charbonnier2012clumpy, bonnivard2016clumpy, 2019CoPhC.235..336H}. We draw 200 realisations from the posterior distributions of the parameters characterising the NFW profile \citep{NFW+1996, Navarro:1996gj} adopted for the MW: \emph{(i)} distance of the Sun to the Galactic centre, $R_{\odot}$, \emph{(ii)} scale radius $r_s$ of the NFW profile, \emph{(iii)} virial radius $r_{200}$ of the MW and \emph{(iv)} the DM density $\rho_{\odot}$ at the position of the Sun.

\begin{figure*}
\begin{centering}
\includegraphics[width=0.98\linewidth]{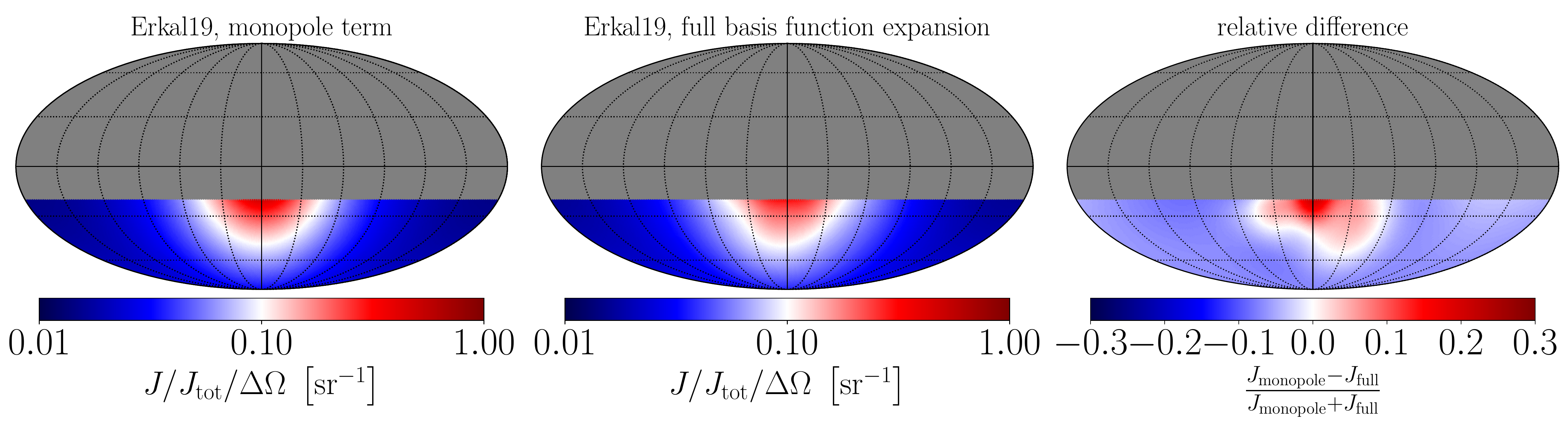}
\caption{Comparison of the two-dimensional MW \Jf-factor profile derived from the Erkal19 simulation either showing only the monopole term of the BFE (left panel) or a MW halo featuring the dynamical response due to the LMC passage according to the full BFE (central panel). Both maps have been normalised by division with the MW's total \Jf-factor for each respective case.  The restriction to Galactic latitudes $b \leq -20^{\circ}$ has been applied to reflect the sky fraction that is optimal for the \Fermi-LAT data analysis presented in the remainder of this work (see Sec.~\ref{sec:fitting_opti}). The right panel illustrates the relative differences between both quantities describing the MW DM halo shape.
\label{fig:jfactor_comparison}}
\par\end{centering}
\end{figure*}

To illustrate the expected deformations induced by the MW-LMC dynamics, we provide a direct comparison in terms of the Erkal19 simulation suite in Fig.~\ref{fig:jfactor_comparison}. The panels display the outer MW halo in the southern hemisphere of the sky at $b \leq -20^{\circ}$, which turns out be the optimal part of the sky to perform the \Fermi-LAT gamma-ray analysis (see Sec.~\ref{sec:fitting_opti}). The left panel shows the  \Jf-factor map associated to the monopole term of the BFE, which has been used to decompose the DM density distribution in the MW, whereas the central panel displays the \Jf-factor map of the MW halo derived from the full BFE of the Erkal19 simulation. Confronting the monopole contribution with the full BFE yields the most faithful assessment of the expected deformations since the initial conditions are the same for both scenarios.
To better highlight the differences, we show the relative ratio of the two quantities in the right panel.

\subsubsection{\Jf-factor all-sky maps for the Large Magellanic Cloud}
\label{sec:jfactor_maps_LMC}

To study the impact of the MW-LMC dynamics on the prospects of indirect searches for DM in the LMC, we repeat the approach outlined in Sec.~\ref{sec:jfactor_maps} for the LMC halo, which results in all-sky maps of the LMC \Jf-factors including the full BFE or only its monopole term for the five simulation suites. Although the DM density $\rho_{\chi}$ has been extracted at a time slice that corresponds to the current stage of the evolution of the MW-LMC system, the position of the LMC differs between the simulations. Moreover, it is not necessarily aligned with the astronomically determined centre of this galaxy, which itself is debated in the literature \citep{vanderMarel:2002kq, 1998ApJ...503..674K, vanderMarel:2013jza}. For definiteness, we define the LMC centre to be located at $\left(\ell, b\right) = \left(280.54^{\circ}, -32.51^{\circ}\right)$ which is the favoured rotational centre derived from stellar kinematics \citep{vanderMarel:2002kq}. All HEALPix \Jf-factor maps are rotated such that the pixel with the largest value coincides with this position thus aligning the DM halo with the stellar centre of the LMC.

To enable a comparison with the expectations from a static LMC DM halo profile, we adopt the parametrization of the NFW and Hernquist profile from Tab.~1 in \cite{Regis:2021glv}. The authors of the latter study have examined the radio emission from the centre of the LMC, and obtained constraints on thermally produced WIMP DM under the assumption of different LMC halo shapes whose parameters have been determined via a fit to rotation curve data \citep{1998ApJ...503..674K}, i.e.~the inner parts of the LMC. We derive all-sky \Jf-factor maps from these two static profiles (NFW and Hernquist) with \textsc{CLUMPY}\footnote{Since the virial radius of the LMC is larger than the distance of the Solar System to this object, we use the `galactic' mode of \textsc{CLUMPY} by defining the LMC centre as the new reference Galactic centre.}. The thereby generated all-sky \Jf-factor maps are by default centred on the Galactic centre. We thus rotate the \textsc{CLUMPY} output on the stellar centre of the LMC as before.

\begin{figure*}
\begin{centering}
\includegraphics[width=0.98\linewidth]{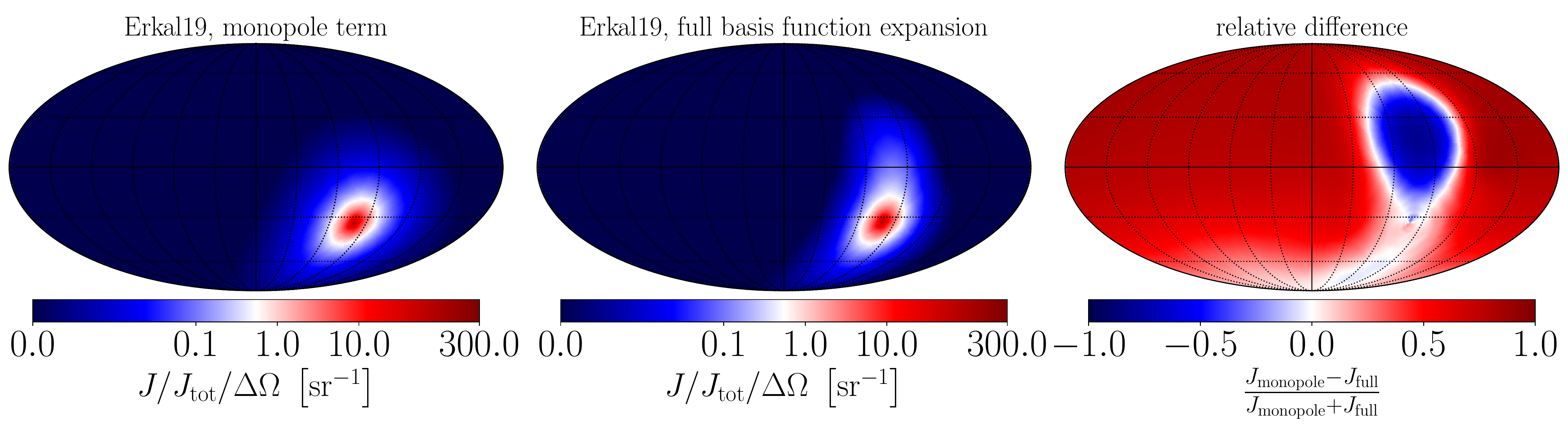}
\caption{Comparison of the two-dimensional LMC \Jf-factor profile derived from the Erkal19 simulation either showing only the monopole term of the BFE (left panel) or the dynamical response due to the LMC passage according to the full BFE (central panel). Both maps have been normalised by division with the respective total LMC \Jf-factor. The right panel illustrates the relative differences between both models of the LMC DM halo shape. 
\label{fig:jfactor_comparison_LMC}}
\par\end{centering}
\end{figure*}

In full analogy to Fig.~\ref{fig:jfactor_comparison}, we visualise the degree of deformations of the LMC DM halo profiles via the Erkal19 simulation in Fig.~\ref{fig:jfactor_comparison_LMC}. The left panel displays the LMC \Jf-factor map taking only into account the monopole term of the BFE whereas the central panel shows a dynamically perturbed LMC halo according to the full BFE of the Erkal19 simulation. The right panel complements both profiles by highlighting the relative differences between the selected cases. Especially the latter panel illustrates the forward wake in the northern hemisphere induced by the LMC passage. 

In comparison to spherically symmetric, static LMC density profiles fitted to rotation curve data -- as done, e.g., in \cite{Regis:2021glv} -- we find that the central part of the LMC is predicted less dense. We stress that the reduced central \Jf-factor is related to the initial conditions of the simulation: the LMC models in both \cite{2022arXiv220501688L} and \cite{Donaldson+2022} are dark matter only and do not include a stellar component for the LMC, which will affect the central dark matter distribution. The halo parameters in \cite{Regis:2021glv} have been fixed via fits to stellar rotation curves and thus incorporate information about the small-scale behaviour around the centre of the LMC while the initial LMC haloes in both \cite{2022arXiv220501688L} and \cite{Donaldson+2022} aim to match the  enclosed LMC mass at larger radii, $\gtrsim 9$ kpc. Since the total mass is rather insensitive to the innermost part of a DM halo, differences between simulated and static LMC profiles may occur.

\subsubsection{Annihilation spectra and gamma-ray flux}
\label{sec:DM_gamma_flux}

In this work, we are referring to a generic thermally produced particle DM candidate at the weak scale, such as those belonging to the class of WIMPs. In addition, we assume that the WIMP candidate is a Majorana fermion ($S_{\chi} = 1$) annihilating into a single Standard Model particle species $f$, i.e.~the branching ratio for this process reads $B_f \equiv 1$. 
We consider a single annihilation channel, namely $\chi\chi \rightarrow b\bar{b}$ to present our results. The associated differential gamma-ray spectrum $\mathrm{d}N_{\gamma}^f/\mathrm{d}E_{\gamma}$, stating the expected average number of photons with energy $E_{\gamma}$ per annihilation event, is taken from \href{http://www.marcocirelli.net/PPPC4DMID.html}{PPPC}~\citep{Cirelli:2010xx}. Eventually, the DM gamma-ray flux depends on two parameters, the DM mass $m_{\chi}$ as well as the velocity-weighted, thermally averaged annihilation cross section  $\langle\sigma v\rangle_\mathrm{ann}$ that determines the intensity of the signal, which is the parameter we ultimately want to constrain.

We stress that the main purpose of our study is to analyse the impact of deformations of the MW halo due to the passage of the LMC on indirect gamma-ray DM searches. These dynamically induced deformations are -- at least for $s$-wave annihilation processes -- merely affecting the spatial morphology of the DM signal, and
their impact is the same no matter what the chosen spectrum is. Thus, the choice of the exact particle DM model plays a minor role on our results and discussion.
Consequently, we focus only on one annihilation channel 
in the main text and provide the results for the
$\tau^+ \tau^-$-channel in Appendix~\ref{app:tau_results}.

\section{\Fermi-LAT data analysis}
\label{sec:fermianalysis}
\subsection{Data selection}
\label{sec:lat_data_selection}

The analysis is based on $\sim$12 years of \Fermi-LAT data (Pass8 reconstruction standard) taken between the 4th August 2008 and the 3rd September 2020. The considered reconstructed gamma-ray energies range from 500 MeV to 500 GeV while we focus on those photons classified as belonging to the \texttt{ULTRACLEANVETO} event class and \texttt{FRONT+BACK} event type. We apply further cuts on the selected sample of photons via the requirement of \texttt{DATA\_QUAL==1 \&\& \\ LAT\_CONFIG==1} and restricting the event zenith angle to $<90^{\circ}$, which reduces the contamination of this sample by Earth limb photons. All work that requires selection, manipulation or simulation of \Fermi-LAT data has been conducted via use of the Fermi Science Tools\footnote{\url{https://github.com/fermi-lat/Fermitools-conda}} (version 2.0.8).

We perform a binned log-likelihood analysis for which we bin the selected data as all-sky maps in the HEALPix format \citep{2005ApJ...622..759G} with $N_{\mathrm{side}}=64$ -- resulting in a mean pixel spacing of $\approx 0.9^{\circ}$ -- as well as 30 logarithmically spaced energy bins. Due to the scarcity of photon events at the highest energies of the LAT's sensitivity range, we rebin the LAT data above 7 GeV as well as all astrophysical background and signal all-sky maps (see  Sec.~\ref{sec:bkg_selection}) into larger macro bins. Hence, the high-energy range is included within the analysis framework by creating the following two macro bins: $\left[7, 20\right]$ GeV and $\left[20, 500\right]$ GeV.\footnote{The generation of these maps respects the initial fine energy binning in order to properly account for the LAT's instrument response functions.}

\subsection{Statistical framework}
\label{sec:stat_framework}

We employ a template-based analysis to constrain DM annihilation in the MW halo, a well-known procedure for \Fermi-LAT gamma-ray analyses. To this end, we seek to describe the processed LAT data map via a set of astrophysical background templates $\{B_j\}_{j\in J}$, which is supposed to capture as best as possible the expected different types of gamma-ray sources in and outside of the MW. To these background components we add the DM signal $S$ and evaluate whether it is preferred by the data and if such a preference is not statistically significant (see Eq.~\ref{eq:TS_discovery} in Appendix \ref{app:signal_recovery}), we set upper limits on the strength of the DM contribution.\\
The foundation of the statistical framework to conduct this kind of analysis is the binned Poisson likelihood function
\begin{equation} 
\label{eq:poisson_L}
\mathcal{L\!}\left(\left.\bm{\mu}\right|\bm{n}\right)=\prod_{i,p}  \frac{\mu_{ip}^{n_{ip}}}{\left(n_{ip}\right)!}e^{-\mu_{ip}} 
\end{equation}
where $p$ denotes the spatial pixels of the all-sky map and $i$ the energy bins of the templates. The linear combination of our set of background and signal templates $\bm{\mu}$ is called \emph{the model} whereas $\bm{n}$ represents the data map to which the model is fitted by maximising the value of the likelihood function. In detail, this linear combination of components is defined as follows:
\begin{equation}
\bm{\mu}=N^{\mathrm{DM}} \bm{S} + \sum_{j\in J}\sum_{i}N_{i}^{B_j}B_{j,i}\rm{,}\label{eq:model_eq}
\end{equation}
where, again, the index $i$ denotes the energy bins used in this analysis. Such a model definition relies on two kinds of normalisation parameters:
\begin{itemize}
    \item Background normalisation parameters $\left\{N_{i}^{B_j}\right\}_{i, j\in J}$ for each background component and energy bin rendering it possible to incorporate spectral fluctuations present in the experimental data and to mitigate potential deviations of the model from reality. Such imperfections of the utilised astrophysical models are expected and thus taken care of. Similar approaches have been employed in the context of template-based analyses, for instance~\cite{TheFermi-LAT:2017vmf, Macias:2019omb}.
    \item A single signal normalisation parameter, $N^{\mathrm{DM}}$, which enables us to exploit both the spectral and spatial shape of the signal component (more details concerning the morphology of the signal are given in Sec.~\ref{sec:jfactor_maps}). 
\end{itemize}

We modify the standard Poisson likelihood function in Eq.~\ref{eq:poisson_L} in two directions: \emph{(i)}, we work with the logarithm of this function, thus turning the statistical inference into a series of function minimizations which are numerically more accessible with well-tested algorithms and software packages; \emph{(ii)}, we employ a pixel weighing scheme to incorporate the impact of instrumental systematic uncertainties. To this end, we adopt the weighted Poisson log-likelihood prescription developed by the \Fermi-LAT collaboration in \cite{Fermi-LAT:2019yla}, which reads
\begin{equation}
\label{eq:weighted_likelihood}
\ln\mathcal{L}_{w}\left(\left.\bm{\mu}\right|\bm{n}\right)=\sum_{i,p}w_{ip}\left(n_{ip}\ln\mu_{ip}-\mu_{ip}\right).
\end{equation}
Per energy bin, each template pixel is assigned a weight $w_{ip}$ whose value we obtain via the Fermi Science Tools calling its routines \texttt{gteffbkg}, \texttt{gtalphabkg} and 
\texttt{gtwtsmap}. These weights are essentially obtained via integration
in space and energy of the provided model or \Fermi-LAT data\footnote{The technical aspects of these routines' implementation is provided at \url{https://fermi.gsfc.nasa.gov/ssc/data/analysis/scitools/weighted_like.pdf} or Appendix B of \cite{Fermi-LAT:2019yla}.}. Since we aim to incorporate the effect of instrumental systematic uncertainties, we rely on a ``data-driven'' weight calculation approach, that is, we utilise the LAT data themselves to derive the weights. Thus, pixels are penalised in particularly bright parts of the gamma-ray sky (point-like/extended sources, diffuse emission along the MW's disc) taking into account potential sources of systematic errors like contamination of the data sample by charged cosmic-ray events, calibration of the instrument's point spread function (PSF) or spectral mismodelling of large-scale diffuse sources. Throughout the analysis, we keep the level of systematic uncertainties to $3\%$ (for all energy bins) following the approach of the \Fermi-LAT collaboration \cite{Fermi-LAT:2019yla}.

Hypothesis testing -- the procedure to discriminate between competing, alternative descriptions of reality, here represented by our model in Eq.~\ref{eq:model_eq} -- is implemented via the log-likelihood ratio test statistic, which in the relevant case of setting upper limits reads
\begin{widetext}
\begin{equation} 
\label{eq:TS_stat_reach} 
\qquad\qquad\qquad\qquad\qquad\qquad\textrm{TS}\!\left(N^{\mathrm{DM}}\right)= \begin{cases} -2\min_{\{N_i^{B_j}\}}\left(\ln\!\left[\frac{\mathcal{L}_w\!\left(\left.\bm{\mu}(N^{\mathrm{DM}},N_i^{B_j}) \right|\bm{n}\right)}{\mathcal{L}_w\!\left(\left.\bm{\hat{\mu}}\right|\bm{n}\right)}\right]\right)\, & N^{\mathrm{DM}} \geq \hat N^{\mathrm{DM}}\\ 0 & N^{\mathrm{DM}} < \hat N^{\mathrm{DM}}\rm{.} 
\end{cases} 
\end{equation}
\end{widetext}
This construction relies on the profiled likelihood function (that is, treating the background normalisations $\left\{N_{i}^{B_j}\right\}_{i, j\in J}$ as nuisance parameters) as discussed in \cite{Cowan:2010js}. Model parameters marked with $\hat{\cdot}$ denote the best-fit values found via minimization of the log-likelihood function. The test statistic in Eq.~(\ref{eq:TS_stat_reach}) only depends on the DM normalisation. Moreover, possible values of $N^{\mathrm{DM}}$ smaller than the best-fit value are discarded, thus, the TS-distribution follows a half-$\chi^{2}$-distribution with one degree of freedom (see Sec.~3.6 of \cite{Cowan:2010js}). Therefore, we set an 95\% confidence level (C.L.) upper limit on $N^{\mathrm{DM}}$ where the test statistic attains a value of 2.71.

All log-likelihood minimization steps are performed with the \textsc{iminuit} python package \cite{iminuit} (version 1.5.3).

\subsection{Astrophysical background model selection}
\label{sec:bkg_selection}

The gamma-ray sky as seen by the LAT can be decomposed into a combination of a multitude of distinct contributions, which either originate in Galactic or extragalactic sources. To constrain the extended signal such as gamma rays from DM annihilation in the outer MW halo, we incorporate the same  astrophysical contributions considered in~\cite{Calore:2021hhn}.
We summarise the components below and refer the interested reader to Appendix~\ref{app:astro_comp} for more details.
\begin{itemize}
\item The interstellar emission (IE) originates from
cosmic-ray interactions with gas and low-energy ambient 
photon fields. Among different IE models and based on 
the findings of~\cite{Calore:2021hhn}, we consider as a baseline IE model the henceforth called \emph{Lorimer I}, taken from the set of realisations considered in the ``1st Fermi LAT Supernova Remnant Catalog''\footnote{The model files have been made public by the \Fermi-LAT collaboration at: \url{https://fermi.gsfc.nasa.gov/ssc/data/access/lat/1st_SNR_catalog/}.} \citep{Acero:2015prw}. Three alternative models utilised in \cite{Calore:2021hhn} -- called \emph{foreground model A, B} and \emph{C} -- are adopted from \cite{Ackermann:2014usa}. More details about the IE models are given in Appendix~\ref{app:astro_comp}.
\item The  isotropic diffuse gamma-ray background (IGRB)
is a large-scale contribution to the 
gamma-ray sky which is spatially isotropic and believed to originate from the 
superposition of many, sub-threshold, sources~\citep{Fornasa:2015qua}.
\item The modelling of the resolved point-like and extended sources
is based on a 10-year data
set, i.e.~the so-called 4FGL-DR2 \citep{Fermi-LAT:2019yla,2020arXiv200511208B}.
\item We also model other large-scale extended gamma-ray emissions from the Fermi Bubbles (FB), Loop I, the Sun and the Moon, following standard practice in LAT data analyses.
\end{itemize}

Passing from these background (and signal) models to templates containing the expected photon counts per sky direction is achieved by dedicated routines of the Fermi Science Tools. For our statistical tests, we generate ``infinite statistics'' or \emph{Asimov} realisations \citep{Cowan:2010js} of the background and signal models via \textit{gtmodel}, which internally processes the full convolution of the input model files with the LAT's instrument response functions associated with the selected gamma-ray data set (c.f.~\ref{sec:lat_data_selection}). We include the LAT's energy dispersion for all background and signal components by adding the parser argument \texttt{edisp\_bins=-1}.

\subsection{Fitting procedure and region-of-interest optimization}
\label{sec:fitting_opti}

We employ the following general analysis rundown and reasoning to statistically soundly and robustly assess the implications on DM pair-annihilation in the MW halo from \Fermi-LAT data:
\begin{enumerate}
    \item Generating a \emph{baseline gamma-ray sky model} from a fit of the full set of astrophysical background templates to the all-sky data.
    \item Including the signal component, preparing the MW halo analysis by shrinking the total region of interest (ROI) to a smaller fraction of the sky that yields a good agreement between the TS-distribution (Eq.~\ref{eq:TS_stat_reach}) with respect to LAT data and baseline model as input data $\bm{n}$. 
    \item Setting upper limits on the DM pair-annihilation cross section  with respect to the optimised ROI and a particular scenario of signal templates.
\end{enumerate}

\subsection*{Deriving a baseline model of the gamma-ray sky}
The importance of a baseline model entirely composed of the background templates considered in this analysis lies in its utility for all future statistical inference. Such a model may be used as an alternative data map that is guaranteed to contain only known gamma-ray emitters, which is not necessarily true for the experimentally observed gamma-ray data. Hence, whenever the TS-distribution as a function of the signal normalisation (Eq.~\ref{eq:TS_stat_reach}) shows a comparable behaviour with respect to baseline model and real LAT data, the selected sky region can reliably be described via the set of background and signal templates at hand. Since the baseline model is a combination of ``infinite statistics'' templates, we can draw Poisson realisations to quantify the expected statistical scatter of the eventually reported upper limits.

Since we employ the same astrophysical gamma-ray emission components as in \cite{Calore:2021hhn}, we repeat the prescription presented therein to derive the baseline model. In short, the algorithm consists of an iterative fitting scheme that splits the entire sky into three disjoint parts defined as follows: (a) \textit{high-latitude} -- $|b|>30^{\circ}$ and neglecting the ``patch''-region (c.f.~\citealt{Fermi-LAT:2019yla}), which is located at $-105^{\circ}\leq\ell\leq60^{\circ}$, (b) \textit{outer galaxy} -- $|b|\leq30^{\circ}$, $|\ell|\geq90^{\circ}$ and (c) \textit{inner galaxy} -- $|b|\leq30^{\circ}$, $|\ell|\leq90^{\circ}$. The  normalisation constants $N_i^{B_j}$ of a particular gamma-ray emitter are only fit in the part of the sky where it mainly contributes to while these parameters are held fixed at the thereby obtained best-fit values when the other regions are addressed. The exact details about the assignment of sky regions to particular background components are provided in Sec.~IV B of \cite{Calore:2021hhn}.

We run the iterative fit for 100 times to eventually derive a baseline fit to the all-sky LAT data. Besides this general scheme, there are a few technical modifications necessary to incorporate the IE models foreground model A, B and C -- henceforth abbreviated as FGMA, FGMB and FGMC -- into this framework as well as the wealth of sources in 4FGL-DR2, which deviates from the original recipe in \cite{Calore:2021hhn}.\\
\emph{Treatment of IE models:} All five IE models feature a single template quantifying the gamma-ray emission from inverse Compton (IC) scattering processes, which we split into three sub-templates whose boundaries coincide with the definitions of the `high-latitude', `outer galaxy' and `inner galaxy' regions of the iterative fit. The same procedure is applied to the gamma-ray emission associated with the gas maps in FGMA-C. After the iterative fit, all IE-related components are multiplied by their best-fit normalisation parameters and added to form an optimised IE template that is utilised in the subsequent analysis parts.\\
\emph{Treatment of 4FGL-DR2 sources:} Combining all sources listed in 4FGL-DR2 into a single template causes the brightest sources in the template to drive the best-fit value of the template's normalisation parameter. We weaken the impact of bright sources by separating sources with an energy flux of $E_{100} < 4\cdot10^{-10}\;\left[\mathrm{MeV}\,\mathrm{cm}^{-2}\,\mathrm{s}^{-1}\right]$ (integrated from 100 MeV to 100 GeV) from those above this threshold. A source that surpasses this threshold is fitted individually during the all-sky fit (d) of an iteration step. We list these bright sources and some of their properties in Tab.~\ref{tab:bright_4fgl} in App.~\ref{app:bright_4FGL}. All others are combined in a single 4FGL template. As for the IE templates, all 4FGL-DR2 templates are combined after the fit according to their best-fit values, hence creating an optimised total 4FGL template to be used in all steps that follow.

\subsection*{Optimising the analysis region of interest}

The strategy to perform an ROI optimization before conducting any statistical inference is heavily based on the approach presented in \cite{Zechlin:2017uzo}. In order to search for an ROI that yields statistically reliable upper limits on the DM annihilation cross section, we resort to the southern hemisphere -- thereby circumventing most of LoopI's contribution to the gamma-ray sky -- and investigate the distribution of the test statistic in Eq.~\ref{eq:TS_stat_reach} for both the true LAT data and the baseline model as data input vector $\bm{n}$.

As the signal spectrum depends on the chosen DM parameters and annihilation channel, we construct a model-independent DM signal -- that still exhibits the currently assumed spatial morphology of the MW halo -- by exchanging the generic WIMP spectra with a power law of spectral index -2. Hence, the signal's spectrum is featureless and yields non-zero photon counts throughout the entire energy range considered in this work. The initial normalisation of the power law is somewhat arbitrary since there is no connection to a physical DM model. We hence choose the normalisation such that the expected counts in the first energy bin are maximally of order unity per pixel. 

The optimisation is carried out by systematically shrinking the ROI boundaries from $\ell\in\left[-180^{\circ},180^{\circ}\right]$
to $\ell\in\left[-90,90^{\circ}\right]$ (symmetrically) with $b\in\left[-90^{\circ},-20^{\circ}\right]$. In addition, we mask the FBs by setting all pixels to zero whose counterpart in the FB template predicts non-zero counts. We introduce the requirement of $b\geq20^{\circ}$ to reduce the impact of IE along the Galactic disc. For each ROI, we scan the TS-distribution with respect to the LAT data and baseline model for $\mathrm{TS}\in\left[0, 25\right]$ as a function of $N^{\mathrm{DM}}$ and use the $\ell^2$-metric to quantify their mutual compatibility. We select those Galactic longitude and latitude ROI boundaries for which this metric is minimal. We stress that this optimisation step must be repeated for each of the five considered IE models. 

\emph{Treatment of the optimised 4FGL-DR2 template:} A special note concerns the treatment of 4FGL catalogue sources. In contrast to the iterative fit, we now mask a circular region around the central position of each source. The mask radius is energy-dependent and corresponds to the $95\%$ containment radius of the LAT's PSF\footnote{See \url{https://www.slac.stanford.edu/exp/glast/groups/canda/lat_Performance.htm} for details.} for the chosen LAT event class and type. For the first three energy bins, however, we reduced the mask radius to $90\%$ of the PSF size. The latter exception has been introduced to ensure a reasonable number of pixels to be non-zero so that a template-based analysis remains feasible.  

We illustrate the TS-distribution comparison in Fig.~\ref{fig:McMillan_example_ROI} of Sec.~\ref{sec:ULIM_McMillan2017} for a particular combination of spatial DM distribution and IE model. In what follows, we will always report the selected optimal ROI. Moreover, we test our analysis pipeline in terms of its capability to recover an injected signal in simulated data. The results of this sanity check are described and discussed in Appendix \ref{app:signal_recovery}.

\subsection{Dedicated analysis of the Large Magellanic Cloud region}
\label{sec:lat_analysis_LMCroi}

Since the LMC passage through the MW halo does not only induce a response in the latter but also in the LMC DM halo itself, we aim to analyse the surroundings of the LMC in a dedicated gamma-ray study. To this end, we utilise the same LAT data selected and described in Sec.~\ref{sec:lat_data_selection} but restrict the ROI to a maximal size of $30^{\circ}\times30^{\circ}$ centred on the stellar position of the LMC at $(\ell, b) = \left(280.54^{\circ}, -32.51^{\circ}\right)$ in agreement with the centre of the LMC \Jf-factor maps discussed in Sec.~\ref{sec:jfactor_maps_LMC}. The pixel size of the binned data is set to $0.1^{\circ}\times0.1^{\circ}$ while the energy binning is adopted from the all-sky data set.

The general analysis and fitting strategy is completely analogous to the scheme outlined in the previous section, with the addition of emission model components of the LMC itself.
A decisive characteristic of the dedicated LMC analysis is the need for an additional astrophysical background component quantifying the expected gamma-ray emission due to cosmic-ray interactions with gas and radiation fields in the LMC. To this end, we include four separate templates adopted from the set of extended templates being part of the 4FGL-DR2 catalogue. The components LMC-Galaxy, LMC-North, LMC-FarWest and LMC-30DorWest, representing the cosmic-ray induced gamma-ray emission of the LMC, have been derived in a previous study of the \Fermi-LAT collaboration \citep{Fermi-LAT:2015bpm} based on a six-year data set. These models are the result of a convolution of gas column density maps of the LMC reported in \cite{2012ApJ...756....5L} and a data-driven approach to extract regions of significant extended gamma-ray emission associated with the LMC. Due to their data-driven nature, these models may already incorporate a contribution from an exotic gamma-ray emitter like DM if taken at face value. The authors of \cite{Buckley:2015doa} comment on this point by analysing the correlation between the DM component and the LMC-related templates. They find a particularly sizeable correlation with the ``LMC-Galaxy'' template, which is centred on the stellar position of the LMC and, at the same time, the most extended among the considered ones. Such a correlation might artificially boost the constraining power of our template-based approach. We mitigate this effect by including the four LMC templates as unmasked, independent model components whose normalisation parameters are a subset of all the model's nuisance parameters that are profiled over to set upper limits on the DM annihilation cross section. Moreover, our ROI is considerably larger than the $10^{\circ}\times10^{\circ}$ ROI adopted in \cite{Fermi-LAT:2015bpm} and \cite{Buckley:2015doa}. Consequently, the spatial extension of the DM component beyond the size of the four LMC templates partially brakes the degeneracy with these templates. 
We anticipate that the later derived upper limits that we find are indeed comparable to the results reported in \cite{Buckley:2015doa}.

The remaining differences to the scheme in Sec.~\ref{sec:fitting_opti} occur in the derivation of a baseline fit for a particular IE model and the optimisation of the analysis ROI. In details, these changes are:
\begin{itemize}
    \item[(a)] All additional point-like gamma-ray sources within $3^{\circ}$ of the stellar position of the LMC are fitted separately by means of a template for each individual source.
    \item[(b)] All detected gamma-ray sources, which already were reported in the 3FGL catalogue \citep{Fermi-LAT:2015bhf} and which are at a distance of $3^{\circ} < r \leq 40^{\circ}$ from the stellar centre of the LMC are cast into a single 3FGL template.
    \item[(c)] All remaining 4FGL-DR2 point-like sources that are neither in 3FGL nor within $3^{\circ}$ of the ROI centre, are cast into another template.
    \item[(d)] Regarding the IE contribution, we consider the following models: We adopt FGMA as baseline model but keep Lorimer I as an alternative test case. This change is motivated by the improved performance of FGMA in the LMC ROI compared to Lorimer I. Regarding the latter, we only fit its IC, ring 3 CO component as well as the HI templates of ring 2 to 4 -- the emission associated to the remaining ones is almost zero in the LMC ROI. FGMA and FGMC are treated as outlined in Sec.~\ref{sec:fitting_opti}. Lastly, we add the Galactic diffuse background model of the \Fermi-LAT collaboration in our list of viable IE models. The reasons for excluding this particular model from the MW halo study do not apply to the LMC region. For example, the data-driven nature of this model is not hampering our efforts because the relevant parts of the sky do not fall into the chosen LMC ROI.
    \item[(e)] The FBs as well as the gamma-ray emission from the Sun and Moon are neglected because their contribution is almost zero in the chosen ROI.
    \item[(f)] All astrophysical background components are fitted in the full ROI at the same time without specifying disjoint fit regions to derive a baseline fit to the gamma-ray data in the ROI.
    \item[(g)] After the baseline fit, the IE components are summed with their best-fit normalisation to an optimised IE model. The localised point-like sources in 4FGL-DR2 are treated analogously except for the LMC-related templates, which we keep as individual templates even in the stage for setting upper limits.
    \item[(h)] The ROI optimisation translates to symmetrically shrinking the width and height of the ROI up to the point where the optimal correspondence between expected TS-distribution and observed TS-distribution (with respect to the auxiliary DM signal template featuring a power law spectrum) is achieved.
    \item[(i)] To explore the TS-distributions and to set upper limits, we mask all detected point-like sources in the same way as outlined before except for the positions of the LMC-related emitters.
\end{itemize}

\section{Constraints on particle dark matter}
\label{sec:upperlimits}

In this section, we present the results of the constraints on DM pair-annihilation processes in the MW and LMC haloes with the analysis framework described in Sec.~\ref{sec:fermianalysis}. As mentioned in Sec.~\ref{sec:simulation}, the characterisation of the gravitational potential of the MW comes with a non-negligible uncertainty. The dynamical impact of the LMC passage may be regarded as a second-order effect that adds to the inherent uncertainties of the available astronomical observations. Consequently, we first explore the expected scatter of constraints on the DM parameter space induced by observational uncertainties of the MW's DM halo while in the following subsections, we shed light on the significance of including the MW's response to the LMC passage for the outcome of indirect searches for DM.

The following results for the MW outer halo have been obtained with our benchmark IE model, Lorimer I, unless stated otherwise, whereas the benchmark for the LMC analysis is FGMA. We investigate the impact of the chosen IE model on the DM constraints in  Appendix~\ref{app:iem_limits} -- we anticipate
that limits at high latitude are mildly affected by the IE choice, and are robust in this respect.

\subsection{Impact of the uncertainty of the Milky Way's gravitational potential}
\label{sec:ULIM_McMillan2017}

We utilise the assessment of the MW's gravitational potential and mass distribution in \citet{2017MNRAS.465...76M} to explore the impact of their uncertainty on DM indirect searches in the outer MW DM halo. To re-iterate, the author of that study assumes a standard NFW profile (inner slope parameter $\gamma = 1$) to describe the MW DM halo and following this fundamental assumption derives, among others, posterior distributions for $ r_s, R_{\odot}, \rho_{\odot}$, and $R_{200}$ to scale and normalise the NFW profile in accordance with the observational constraints. From these posterior distributions we randomly draw 200 points and generate the associated all-sky \Jf-factor map of the MW. For each of these MW realisations, we derive \Fermi-LAT upper limits on the DM pair-annihilation cross section  following the scheme outlined in Sec.~\ref{sec:fitting_opti}.

\begin{figure*}
\centering
\includegraphics[width=0.70\textwidth]{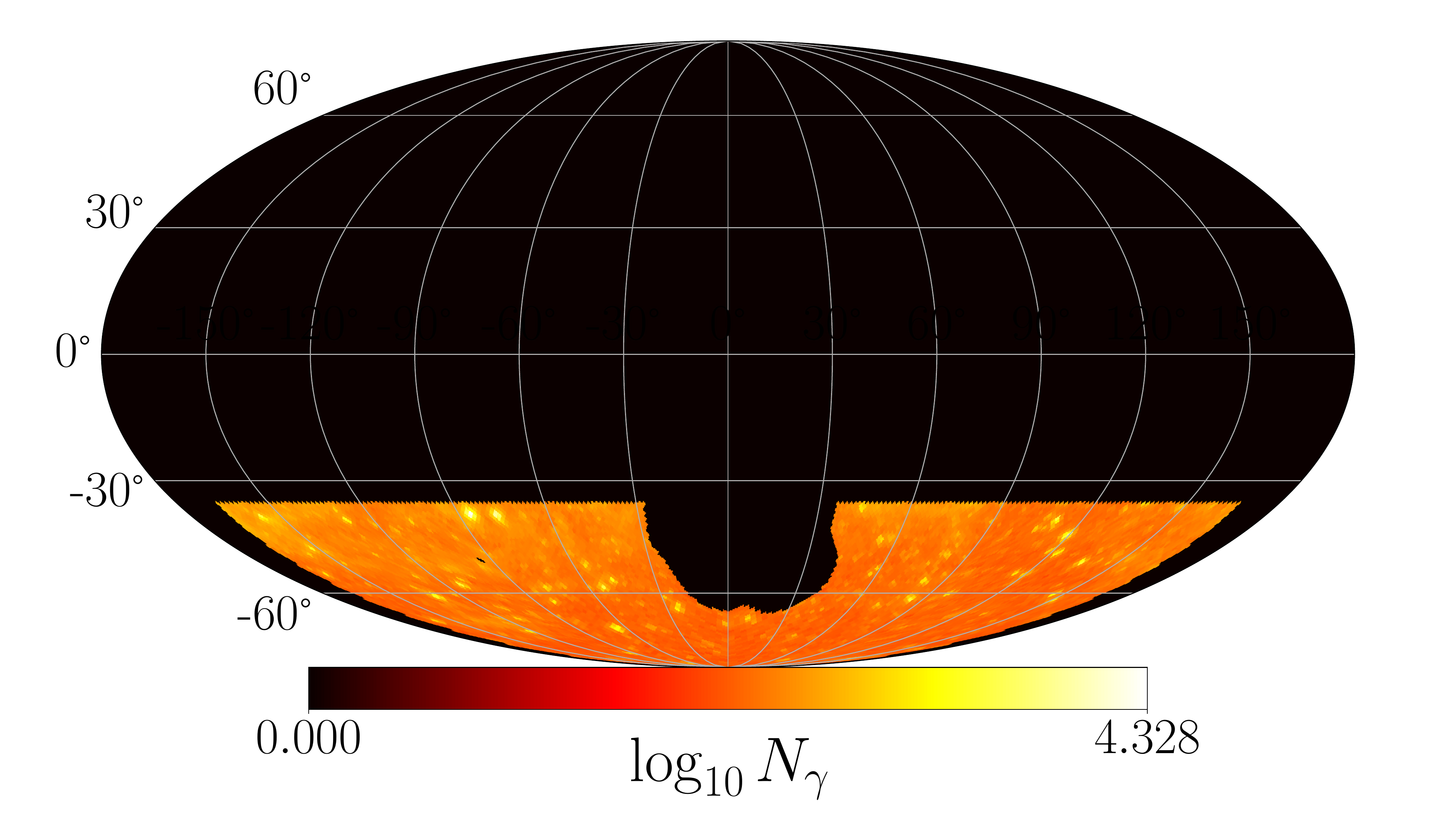}
\caption{\Fermi-LAT gamma-ray data integrated over this analysis energy range (500 MeV to 500 GeV) within the optimal ROI determined in the first step of our statistical framework. It is defined by $\ell\in\left[-167^{\circ}, 167^{\circ}\right]$ and $b\in\left[-90^{\circ}, -35^{\circ}\right]$.
\label{fig:optimal_roi}}
\end{figure*}

\begin{figure*}
\centering
\includegraphics[width=0.45\textwidth]{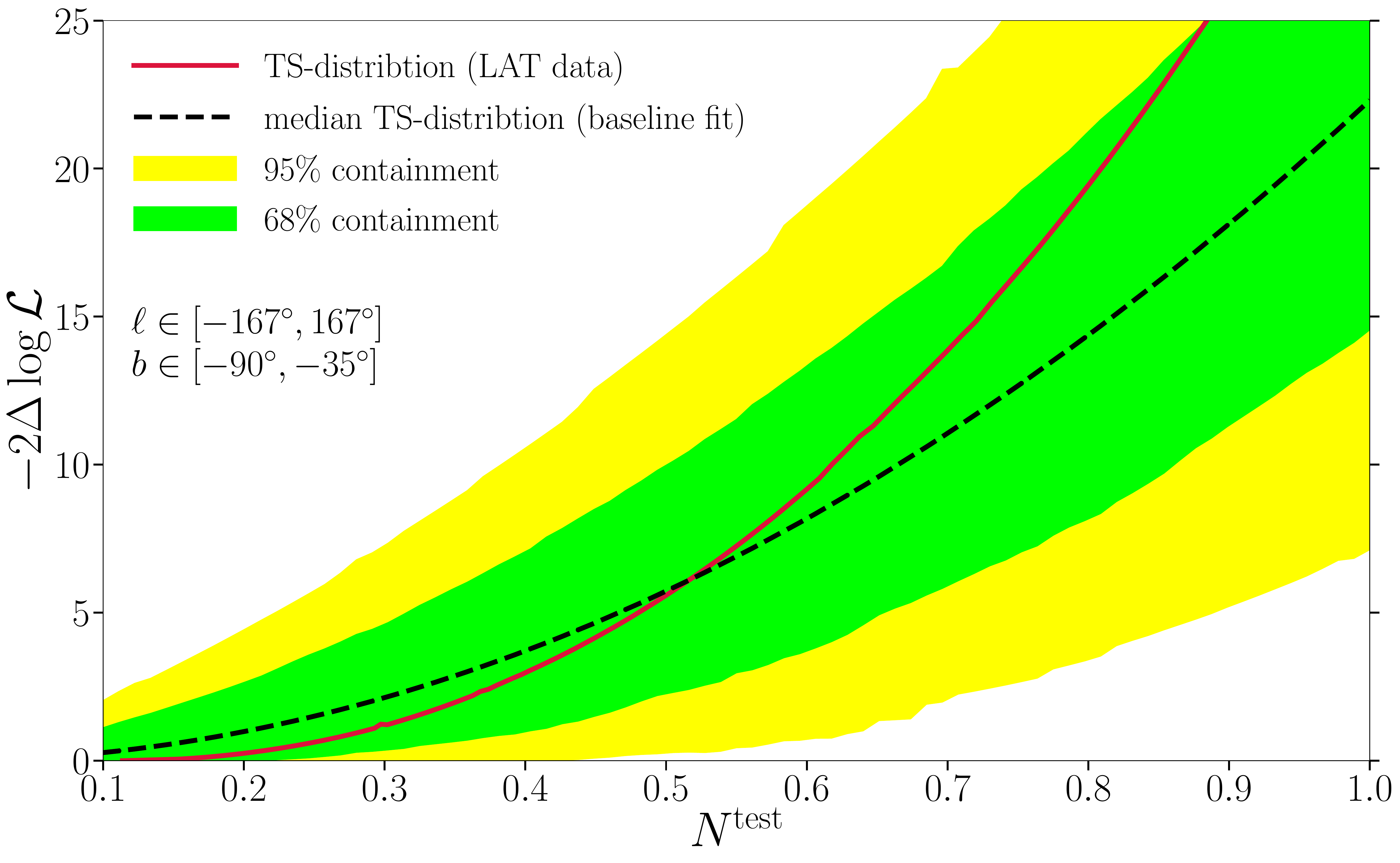}
\hfill
\includegraphics[width=0.52\textwidth]{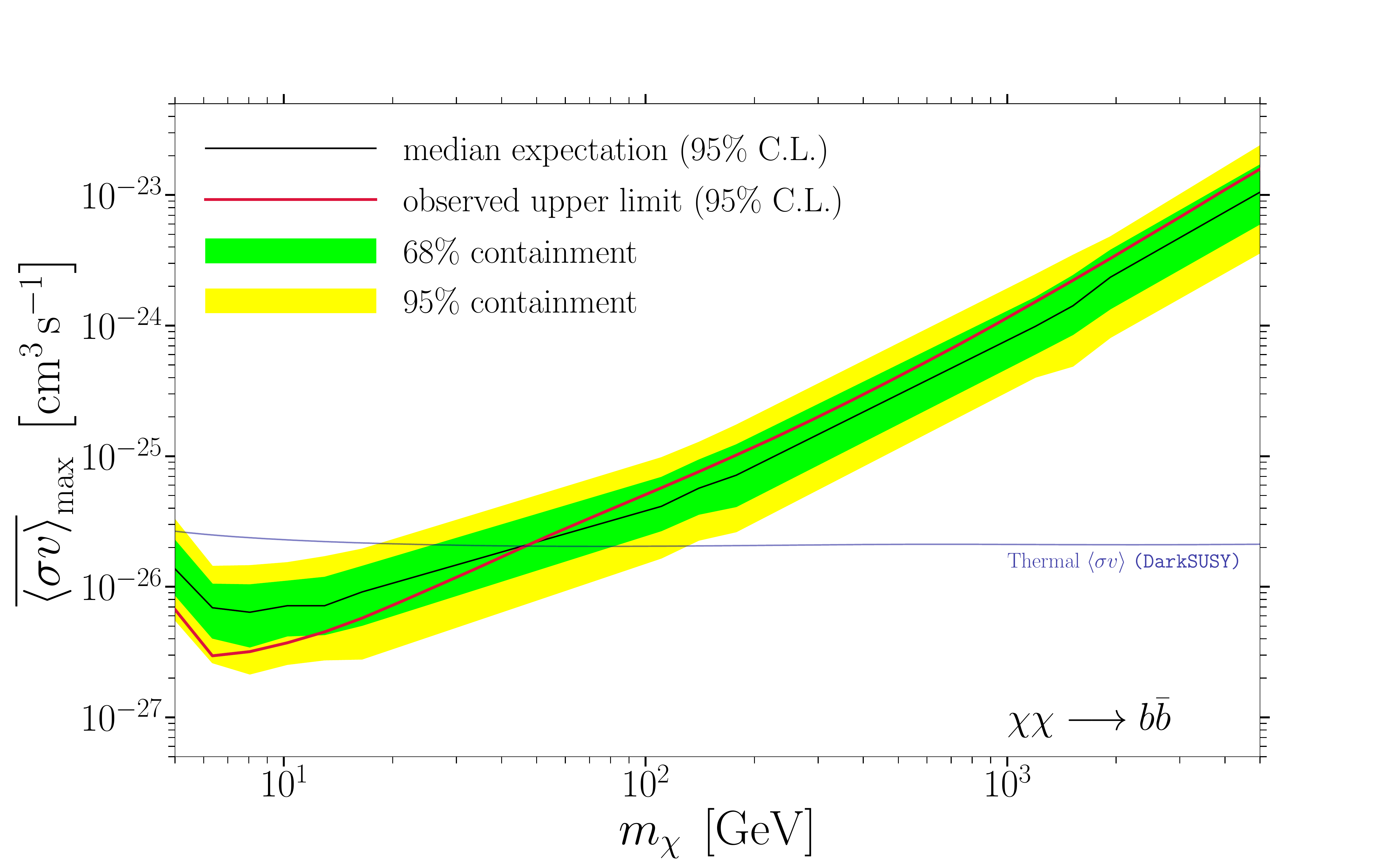}
\caption{Results of the ROI optimisation scheme presented in Sec.~\ref{sec:fitting_opti} for the particular instance of MW halo parameters $\protect(r_s, R_{\odot}, \rho_{\odot},R_{200}) = (12.06\unit{kpc}, 8.30\unit{kpc}, 0.39\unit{GeV}\unit{cm}^{-3},199.03\unit{kpc})$ drawn from the posterior distributions in \protect\cite{2017MNRAS.465...76M}.  (\emph{Left}:) TS-distribution with respect to the baseline fit for IE model Lorimer I in terms of the median expectation and its statistical scatter (green: 68$\%$ containment, yellow: 95$\%$ containment). The red line indicates the observed TS-distribution with respect to the select LAT data in the designated ROI. The normalisation parameter $\protect N^{\mathrm{test}}$ is an auxiliary parameter without physical relevance that controls the strength of the inject DM signal following a power law with spectral index -2. (\emph{Right}:) Median $95\%$ C.L.~upper limits (black) on the velocity-weighted thermally averaged DM pair-annihilation cross section  depending on the assumed DM mass $m_{\chi}$ for the prompt gamma-ray emission from DM annihilating into $b\bar{b}$ final states and its statistical scatter (green: 68$\%$ containment, yellow: 95$\%$ containment) derived with respect to the baseline fit with Lorimer I. The red line displays the respective observed upper limits. The blue line indicates the thermal WIMP annihilation cross section  for a DM particle with mass $m_{\chi}$ to generate the measured cosmological abundance of DM in the early universe (computed with \texttt{DarkSUSY} \protect\citep{Bringmann:2018lay} and current Planck data \protect\citep{Planck:2018vyg}).
\label{fig:McMillan_example_ROI}}
\end{figure*}

As the first step of this analysis pipeline, we search for an optimal ROI in the gamma-ray sky by successively shrinking the considered fraction of the southern hemisphere in Galactic longitude and latitude. We illustrate the statistical performance of the thus obtained optimised ROI via one particular realisation from the McMillan posterior distributions characterised by the tuple of parameters $(r_s, R_{\odot}, \rho_{\odot}, r_{200}) = (12.1\unit{kpc}, 8.3\unit{kpc}, 0.4\unit{GeV}\unit{cm}^{-3},199.0\unit{kpc})$ in Fig.~\ref{fig:McMillan_example_ROI}. We find the best-suited analysis ROI to be defined by $\ell\in\left[-167^{\circ}, 167^{\circ}\right]$ and $b\in\left[-90^{\circ}, -35^{\circ}\right]$, which ensures a reasonable compromise between the constraining power of the remaining gamma-ray sky and the statistical robustness of the resulting upper limits. We show the \Fermi-LAT data integrated over the analysis' energy range inside this optimal ROI in Fig.~\ref{fig:optimal_roi}.

Both panels in Fig.~\ref{fig:McMillan_example_ROI} have been obtained from 200 Poisson realisations of the baseline fit. On one hand, the left panel of this figure shows that the TS-distribution on LAT data stays within the 68$\%$ containment band of the scatter of the baseline fit's TS-distribution. On the other hand, the width of its median's parabola deviates from its analogue on LAT data. This qualitatively indicates that the chosen ROI may contain further gamma-ray emission components which are not or not fully accounted for in the selected set of astrophysical background contributors. Hence, this ROI may feature further emission components that are degenerate with the DM signal. Consequently, the $95\%$ C.L.~upper limits fall within the 68$\%$ containment band derived from mock data for most of the scanned DM masses except for light DM below $\lesssim 10$ GeV. Here, the constraints are slightly stronger (at the $2\sigma$ level) than expected from the baseline fit. We have made sure that the chosen ROI is suitable for all of the 200 random parameter tuple $(r_s, R_{\odot}, \rho_{\odot}, R_{200})$. In fact, the observed moderate fluctuation for light DM is a common feature among all of these realisations of the MW halo. From a qualitative point of view, the consistency between the TS-distribution of mock and real data in this work is similar to the benchmark scenario studied and described in Fig.~1 of \cite{Chang:2018bpt}. There, the authors find the same slight deviation for light DM but they also show a more pronounced deviation for DM around the TeV scale where our TS-distributions differ only at the $1\sigma$ level. However, the selected optimal ROIs in both analyses are largely disjoint since we exclude the position of the FBs. The optimised ROI in \cite{Zechlin:2017uzo} (c.f.~Fig.~5 therein) shows a larger overlap with our analysis ROI. The reported accordance of the statistical expectations from their baseline fit and the corresponding performance on real data in their Fig.~4 is well in line with the results presented in this work, i.e.~consistency at the $1\sigma$ level for most of the tested parameter space. The comparison with both literature results corroborates that our optimal ROI provides statistically sound upper limits from DM annihilation processes in the outer MW halo.

\begin{figure*}
\centering
\includegraphics[width=0.49\linewidth]{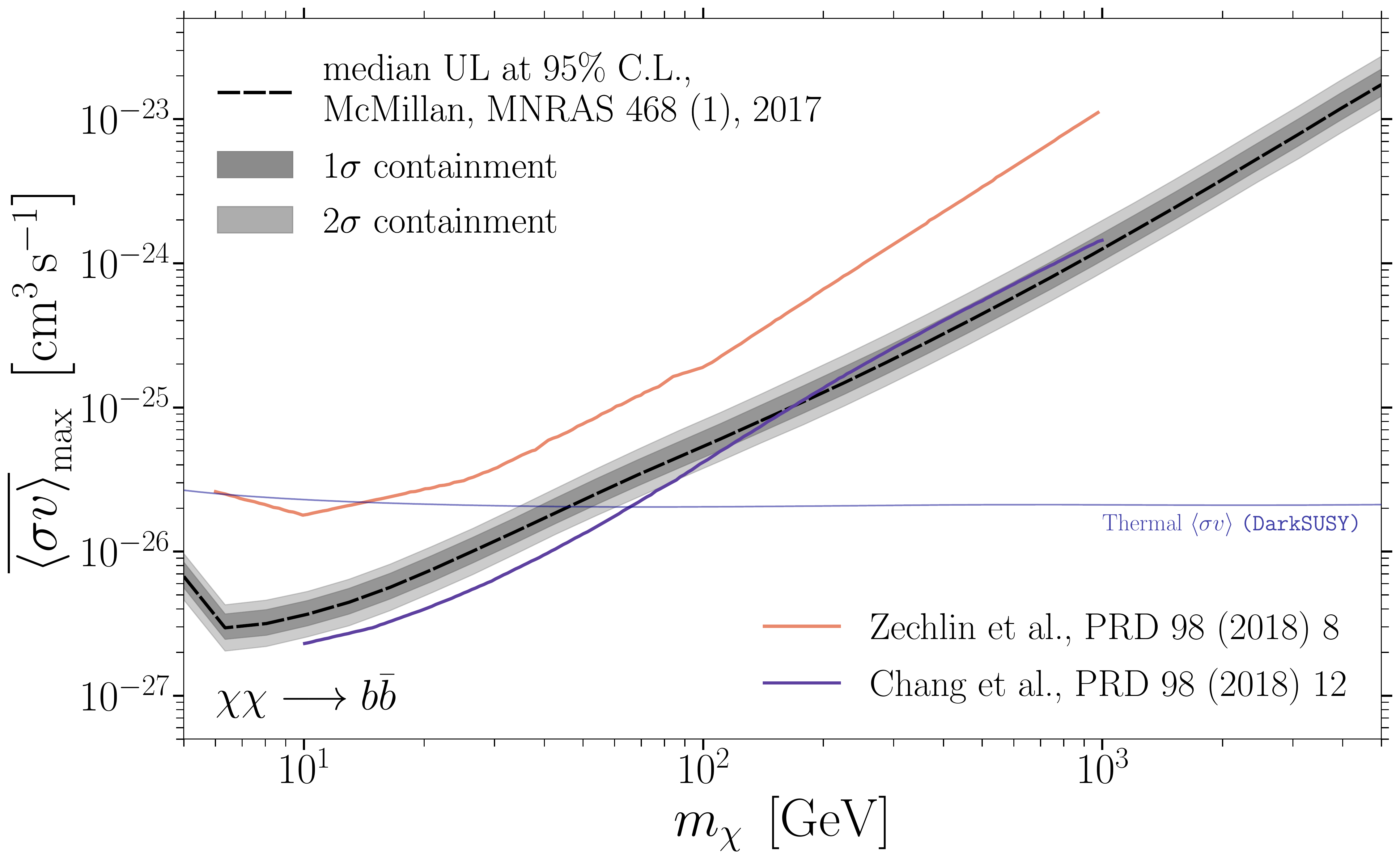}
\includegraphics[width=0.49\linewidth]{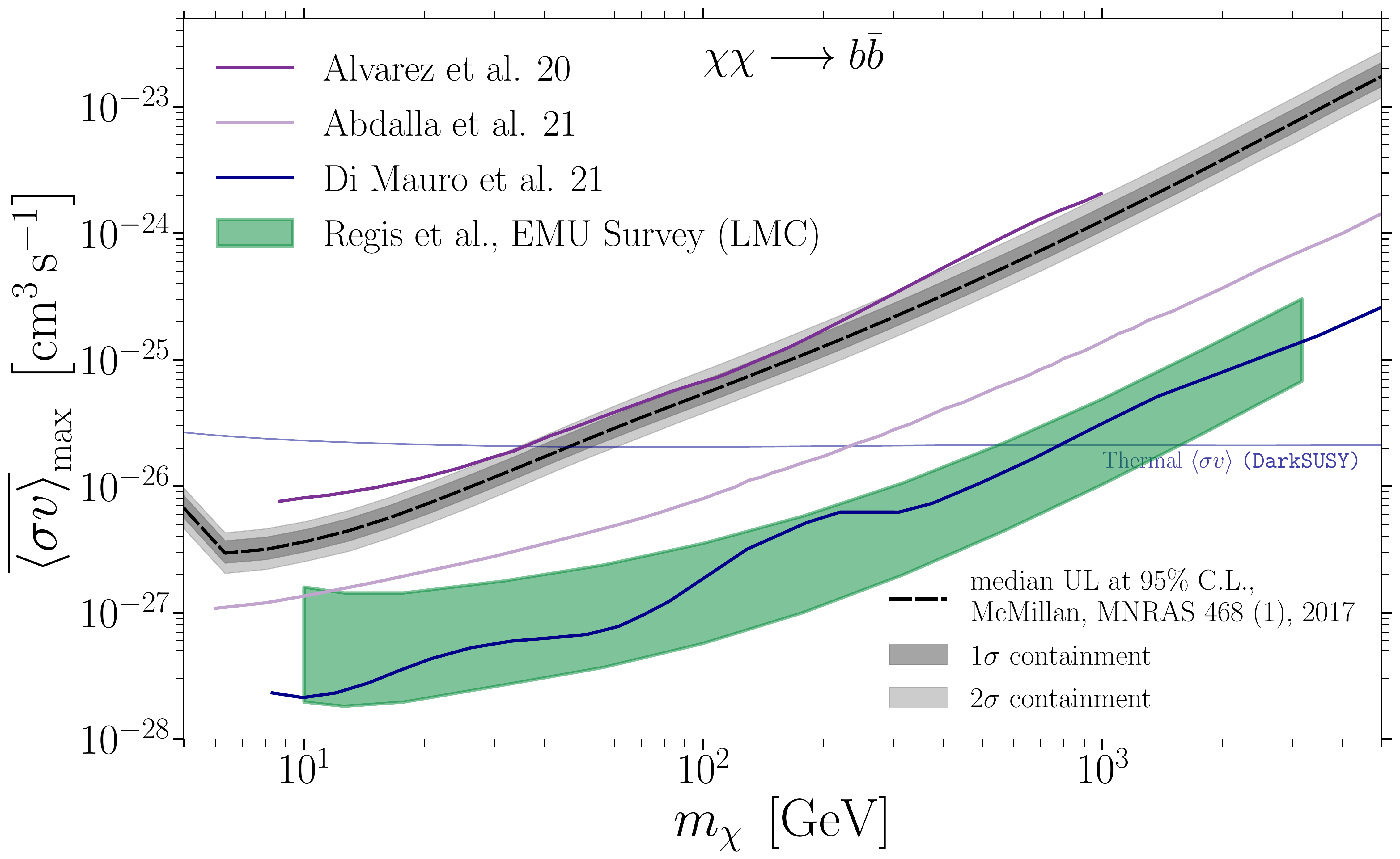}
\caption{Median $95\%$ C.L.~upper limits (dashed black line) on the velocity-weighted thermally averaged DM pair-annihilation cross section  depending on the assumed DM mass $m_{\chi}$ for the prompt gamma-ray emission from DM annihilating into $b\bar{b}$ final states. The upper limits have been derived with respect to our benchmark IE model Lorimer I. The all-sky map in the upper left corner illustrates the optimised size of the analysis ROI of $\ell\in\left[-167^{\circ},167^{\circ}\right]$ and $b\in\left[-90^{\circ},-35^{\circ}\right]$. We obtained the median value from 200 samples randomly drawn from the posterior distributions of the MW DM halo parameters in \protect\cite{2017MNRAS.465...76M} and their associated NFW DM profiles. The dark grey band represents the observed $1\sigma$ scatter of the upper limits while the light grey band denotes the respective $2\sigma$ scatter due to the observational uncertainty of the MW's gravitational potential. (\emph{Left}:) Comparison with independent analyses that derive $95\%$ C.L.~upper limits on DM annihilation in the outer MW halo from high-latitude ($|b| > 20^{\circ}$) \Fermi-LAT data. The orange line refers to the constraints in Fig.~6 of \protect\cite{Zechlin:2017uzo} for the energy band from 1 GeV to 2 GeV while the purple line states the upper limits given in Fig.~1 of \protect\cite{Chang:2018bpt}. (\emph{Right}:) Comparison with indirect multi-messenger probes of thermal DM in terms of $95\%$ C.L.~upper limits on the annihilation cross section. We display bounds from the gamma-ray emission of a set of dwarf spheroidal galaxies obtained in a data-driven approach \protect\citep[][light purple]{Alvarez:2020cmw} as well as dwarf spheroidal galaxies constraints derived with traditional techniques obtained from combining ground-based and space-borne gamma-ray instruments \protect\citep[][dark purple]{Hess:2021cdp}. The dark blue line indicates bounds from AMS-02 $\bar{p}$-data \protect\citep{DiMauro:2021qcf} for the \textsc{med} DM density model that fits the Galactic centre excess while the green band illustrates the range of radio constraints following from the EMU Survey of the LMC \protect\citep{Regis:2021glv}.
\label{fig:McMillan_limits}}
\end{figure*}

The left panel of Fig.~\ref{fig:McMillan_limits} summarises the set of constraints derived from these 200 random realisations of the MW DM halo assuming WIMP DM pair-annihilating into $b\bar{b}$ final states that eventually generate a prompt gamma-ray flux via further processes. The median $95\%$ C.L.~upper limits obtained within the optimised ROI and with respect to LAT data are denoted by a black dashed line whereas the corresponding $1\sigma/2\sigma$ containment bands are depicted as dark grey/light grey shaded bands. The impact of the observational uncertainty of the MW's gravitational potential is less than a factor of two at the $2\sigma$ level across the entire range of DM masses considered in our analysis. This finding may seem astonishing at first glance. It is well known that the uncertainty of the shape of the MW's DM halo has a much larger effect on indirect searches towards the Galactic centre since the DM distribution in this region of the Galaxy can be peaked, core-like or even be completely devoid of DM \citep[see, for example,][]{2011JCAP...11..029I, 2015NatPh..11..245I, 2017PDU....15...90I, 2019JCAP...09..046K, Benito:2020lgu}. However, high Galactic latitudes as inspected by us are dominated by the outer MW DM halo, that is, we probe a much larger volume of the total halo with guaranteed DM presence in order to stabilise the Galactic rotation curve of the MW. Hence, small-scale uncertainties of the MW's gravitational potential, for instance in the Galactic centre, are washed out by the fact that we investigate a large volume of the MW DM halo and probe its cumulative gravitational imprint.

Results for the $\tau^+ \tau^-$ DM annihilation channel are provided in Appendix~\ref{app:tau_results}.

\subsection{Impact of the perturbation of the Milky Way's dark matter halo caused by the Large Magellanic Cloud's passage}
\label{sec:ULIM_MW+LMC}

As discussed in Sec.~\ref{sec:simulation}, the passage of the LMC through the MW halo induces dynamical responses, which add to the already discussed uncertainty of the MW's gravitational potential. The respective responses in the form of wakes are particularly present in the outer MW halo trailing the LMC orbit or developing in front of its current orbital direction. Hence, this dynamical effect is supposed to be detectable in the southern hemisphere, as visualised in Fig.~\ref{fig:jfactor_comparison}.

\begin{figure*}
\centering
\includegraphics[width=0.7\textwidth]{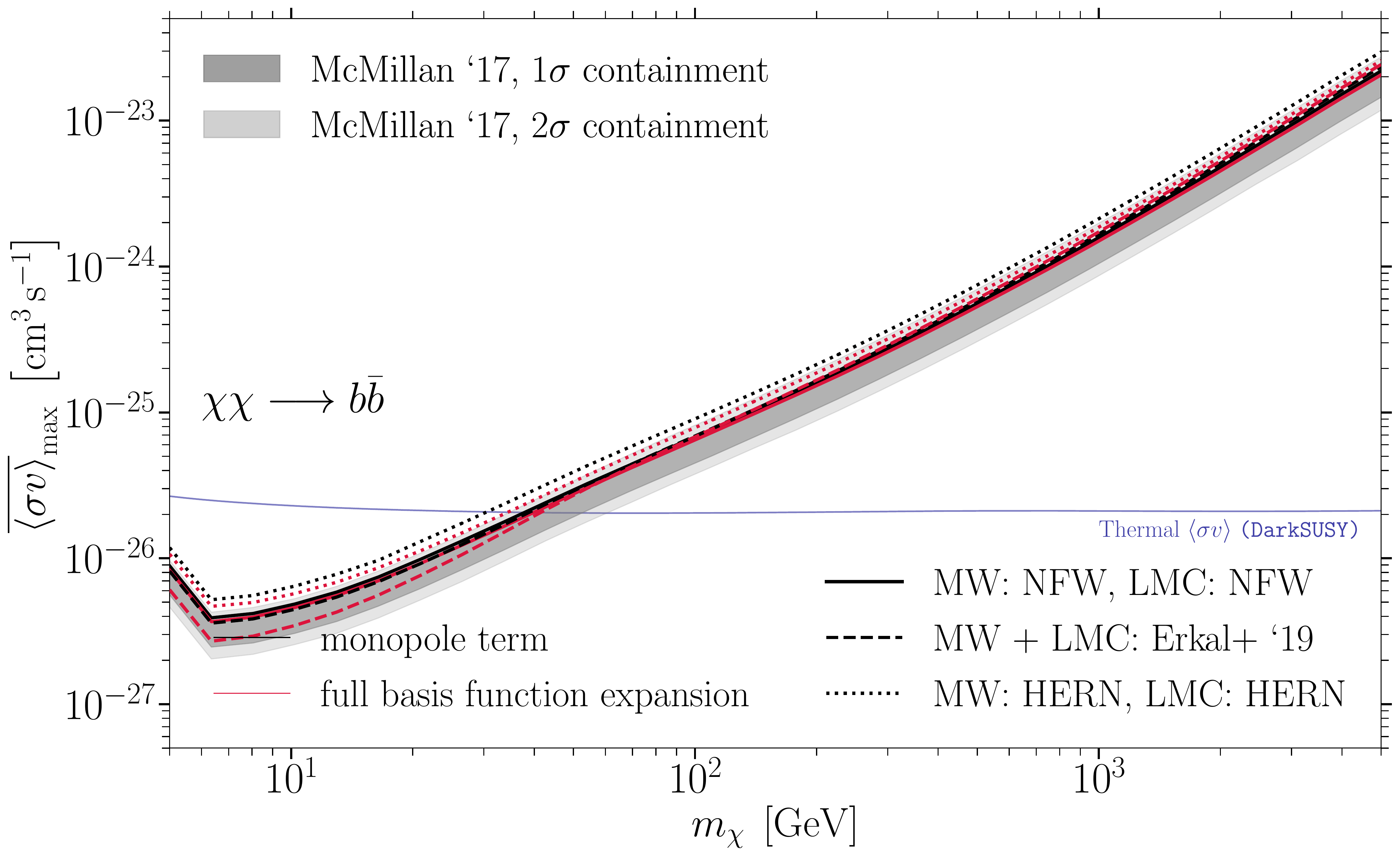}
\caption{$95\%$ C.L.~upper limits on the velocity-weighted thermally averaged DM pair-annihilation cross section  depending on the assumed DM mass $m_{\chi}$ for the prompt gamma-ray emission from DM annihilating into $b\bar{b}$ final states. The black lines illustrate the constraints using only the monopole term to characterise the MW halo whereas the red lines show the corresponding results when the full multipole expansion of the halo is included. The solid (dotted) lines denote the MW halo profile obtained from a set of $N$-body simulations in Ref.~\protect\cite{Donaldson+2022} adopting as initial profiles for the MW an NFW (Hernquist) profile and an NFW (Hernquist) profile for the LMC. The dashed lines represent the halo profiles from the simulation in Ref.~\protect\cite{2019MNRAS.487.2685E, 2020MNRAS.498.5574E}. In each case, the utilised \Jf-factor map is comprised of both the MW and LMC halo. The upper limits have been derived with respect to our benchmark IE model Lorimer I. The all-sky map in the upper left corner illustrates the optimised size of the analysis ROI of $\ell\in\left[-167^{\circ},167^{\circ}\right]$ and $b\in\left[-90^{\circ},-35^{\circ}\right]$. The grey bands have the same meaning as their equivalents in Fig.~\ref{fig:McMillan_limits}. \label{fig:LMC_impact_on_MW}}
\end{figure*}

We quantitatively examine the importance of the LMC passage via the simulations of the MW-LMC system described and discussed in Sec.~\ref{sec:mw-lmc-sim}. To this end, we confront in Fig.~\ref{fig:LMC_impact_on_MW} the $95\%$ C.L.~upper limits on the DM pair-annihilation cross section  ($\chi\chi\longrightarrow b\bar{b}$) obtained from three different simulations with the previously derived uncertainty of the same limits due to observational uncertainty of the MW's gravitational potential. We distinguish constraints for the cases of taking into account solely the MW DM halo's monopole term from the BFE (black lines) and the corresponding full BFE (red lines) according to the respective simulation. As concerns the chosen simulations, we have selected two models from the \citet{Donaldson+2022} suite -- assuming an initial NFW (solid)/HERN (dotted) profile for both the MW and the LMC -- and the Erkal19 simulation (dashed). We have checked that the optimised ROI discussed in Sec.~\ref{sec:ULIM_McMillan2017} can also be applied to these signal DM profiles. We emphasize that the utilised \Jf-factor maps are the sum of both the MW and LMC halo, although the additional boost due to the LMC halo is marginal.

A comparison of the black curves reveals that the different initial conditions in terms of MW halo profile are well within the $2\sigma$ range of the uncertainty reported by \citet{2017MNRAS.465...76M} and thus plausible representations of the physically realised halo of the MW. If we compare the results in red that incorporate the full influence of the LMC as a perturber of the MW halo with those that disregard this effect, we see that the impact on the DM constraints depends on the respective simulation. While both simulations from \cite{Donaldson+2022} predict an almost negligible improvement of the upper limits, the Erkal19 simulation yields a much stronger response that induces an improvement for light DM as large as the $1\sigma$ uncertainty of the MW's gravitational potential. The difference of the obtained bounds is not related to the total \Jf-factor of the full MW halo -- which is the highest for the NFW-NFW simulation -- but rather correlated with the appearance of deviations from the static, spherically symmetric DM halo scenario. Local over- and under-densities induced by the dynamical response of the MW halo are most pronounced in the Erkal19 simulation, which explains the prominent effect in Fig.~\ref{fig:LMC_impact_on_MW} compared to the remaining simulation suites. The difference is thus a manifestation of the initial conditions of each model simulation (i.e. assumptions about extent, profile, mass, internal structure/deformation and components) of the MW-LMC system. We discuss these properties and their interplay in Sec.~\ref{sec:discussion_systematics_MW}. 

\subsection{Indirect searches towards the Large Magellanic Cloud}
\label{sec:ULIM_LMC}

We assess the constraining power of the LMC as a target for indirect DM searches and the importance of including the altered shape (and mass) of the LMC DM halo due to its gravitational interaction with the MW halo. This analysis follows the outline given in Sec.~\ref{sec:lat_analysis_LMCroi}. We stress again that we select FGMA as the benchmark IE model because it shows a better performance than Lorimer I in this particular ROI. We discuss the systematic uncertainty due to the chosen IE model for this dedicated LMC analysis in Appendix \ref{app:iem_limits} and Fig.~\ref{fig:iem_uncertainty_main} therein.

\begin{figure*}
\centering
\includegraphics[width=0.49\textwidth]{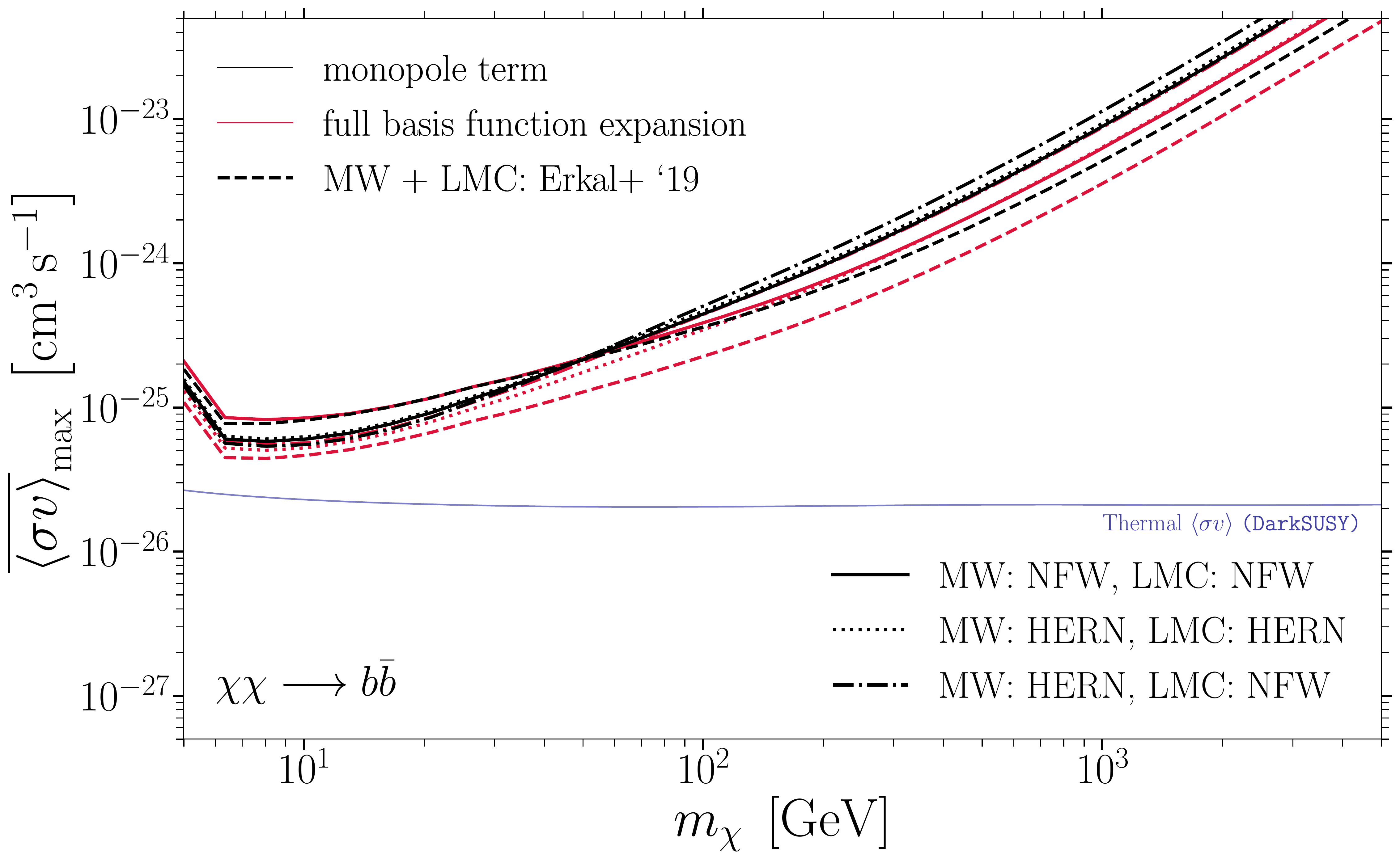}
\hfill
\includegraphics[width=0.49\textwidth]{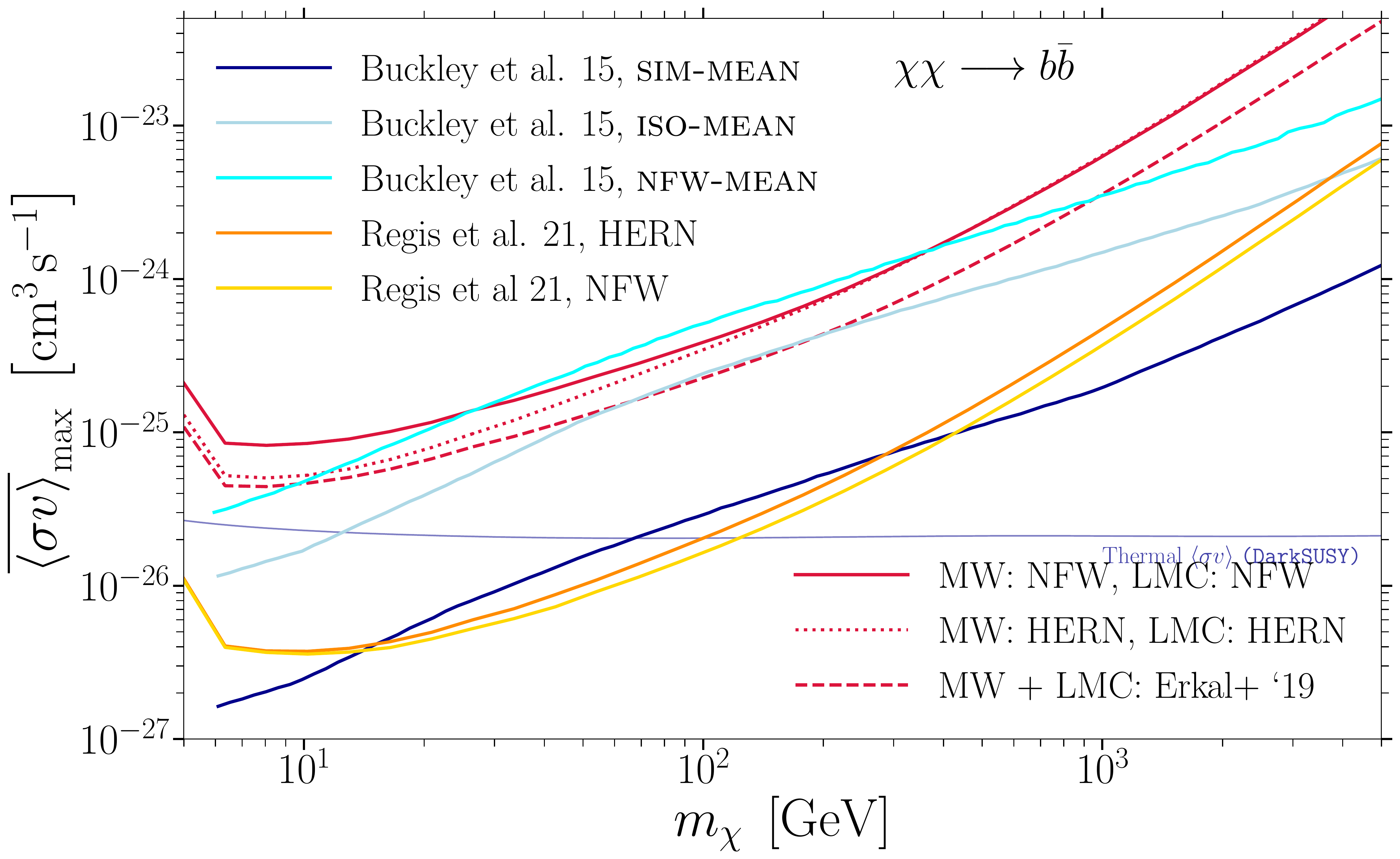}
\caption{$95\%$ C.L.~upper limits on the velocity-weighted thermally averaged DM pair-annihilation cross section  depending on the assumed DM mass $m_{\chi}$ for the prompt gamma-ray emission from DM annihilating into $b\bar{b}$ final states and the LMC DM halo as the target assuming IE according to FGMA. (\emph{Left}:) Impact of the dynamical response of the MW-LMC system illustrated via five simulations of the LMC passage. Black lines represent the monopole term of the LMC halo whereas red lines display the respective scenario including the full BFE. The select ROI size for each case shown in the figure is given in Tab.~\ref{tab:lmc_rois}. (\emph{Right}:) Comparison between the constraints derived from simulated LMC haloes (red) using the full multipole expansion and similar DM halo profiles (orange) adopted from a recent study of the LMC in the radio band \citep{Regis:2021glv} quantifying the morphology of a static LMC. In addition, we provide the bounds derived in an independent gamma-ray analysis \citep{Buckley:2015doa} of the LMC with five years of \Fermi-LAT data for three distinct, static DM halo profiles labelled \textsc{sim-mean} (dark blue), \textsc{iso-mean} (light blue) and \textsc{nfw-mean} (cyan).   \label{fig:LMC_ROI_study}}
\end{figure*}

In Fig.~\ref{fig:LMC_ROI_study}, we summarise our findings regarding the analysis of the LMC's DM halo. In the left panel, we compare the constraints on the DM pair-annihilation cross section  (channel: $\chi\chi\rightarrow b\bar{b}$) using either the full BFE (red) or only its monopole term for the LMC halo (black) of a particular simulation of the evolution of the MW-LMC system. For each of these sets of halo profiles, we optimised the square ROI centred on the LMC position in terms of its size. The optimal ROI sizes are stated in Tab.~\ref{tab:lmc_rois}. We observe a mild impact of the dynamical response of the LMC on the final upper limits for the four simulations from Ref.~\cite{Donaldson+2022} whereas -- and as we had already noticed in the case of the outer MW halo -- the Erkal19 simulation suggests a more pronounced effect, which may be as large as a factor of two. We discuss the observed simulation-dependent variations in more detail in Sec.~\ref{sec:discussion_systematics_LMC}. 

The right panel of Fig.~\ref{fig:LMC_ROI_study} puts the results from the perturbed LMC haloes in the context of the current state-of-the-art in the field. We confront a subset of the DM upper limits from the left panel (red curves) to constraints derived with standard DM haloes following the profiles from  an NFW or Hernquist profile with parameters adopted from~\cite{Regis:2021glv} presented in Sec.~\ref{sec:jfactor_maps_LMC} (orange lines). 
The latter set of upper limits hence represent a static LMC halo that does not incorporate deviations from spherical symmetry. While both types of LMC haloes agree on the strength of the constraints for light DM particles, they differ for larger DM masses. In fact, haloes from~\cite{Regis:2021glv} feature a more pronounced cusp towards the LMC's centre whereas our simulated haloes appear less peaked and smoother in general. The discussion is continued in Sec.~\ref{sec:discussion_systematics_LMC}. 
Representative results of the \Fermi-LAT collaboration’s search for DM in the LMC \citep{Buckley:2015doa} with five years of data are displayed with different shades of blue. We comment on the comparison to our bounds in Sec.~\ref{sec:conclusions}.

\begin{table*}
\begin{centering}
\begin{tabular}{c c c}
\hline 
\multirow{3}{*}{simulation/DM profile} & only monopole & multipole expansion\tabularnewline
\cline{2-3} \cline{3-3} 
 & ROI size & ROI size\tabularnewline
 & \Jf-factor {[}GeV$^2$cm$^{-5}${]} & \Jf-factor {[}GeV$^ 2$cm$^{-5}${]}\tabularnewline
\hline 
\hline
\multirow{2}{*}{Erkal+, `19} & $29.9^{\circ}\times29.9^{\circ}$ & $29.8^{\circ}\times29.8^{\circ}$\tabularnewline
 & $3.39\times10^{20}$ & $3.55\times10^{20}$\tabularnewline
\hline 
\multirow{2}{*}{MW: NFW, LMC: NFW} & $29.9^{\circ}\times29.9^{\circ}$ & $30.0^{\circ}\times30.0^{\circ}$\tabularnewline
 & $1.46\times10^{20}$ & $1.70\times10^{20}$\tabularnewline
\hline 
\multirow{2}{*}{MW: NFW, LMC: HERN} & $29.8^{\circ}\times29.8^{\circ}$ & $29.7^{\circ}\times29.7^{\circ}$\tabularnewline
 & $1.53\times10^{20}$ & $1.83\times10^{20}$\tabularnewline
\hline 
\multirow{2}{*}{MW: HERN, LMC: NFW} & $29.5^{\circ}\times29.5^{\circ}$ & $29.5^{\circ}\times29.5^{\circ}$\tabularnewline
 & $1.45\times10^{20}$ & $1.70\times10^{20}$\tabularnewline
\hline 
\multirow{2}{*}{MW: HERN, LMC: HERN} & $29.8^{\circ}\times29.8^{\circ}$ & $29.7^{\circ}\times29.7^{\circ}$\tabularnewline
 & $1.54\times10^{20}$ & $1.85\times10^{20}$\tabularnewline
\hline 
\multirow{2}{*}{NFW \citep{Regis:2021glv} } & / & $27.5^{\circ}\times27.5^{\circ}$\tabularnewline
 & / & $1.07\times10^{20}$\tabularnewline
\hline 
\multirow{2}{*}{Hernquist \citep{Regis:2021glv}} & / & $27.5^{\circ}\times27.5^{\circ}$\tabularnewline
 & / & $0.98\times10^{20}$\tabularnewline
\hline 
\end{tabular}
\par\end{centering}
\caption{Summary of the optimised ROI sizes for the dedicated study of the LMC environment. The optimisation has been performed under the assumption of the FGMA IE model. For each simulation and basis function expansion scenario we state the total \Jf-factor contained within the reported ROI to facilitate better comparison between the different cases. \label{tab:lmc_rois}}
\end{table*}

\section{Discussion}
\label{sec:discussion}

\subsection{Systematic uncertainties affecting the study of the outer Milky Way halo}
\label{sec:discussion_systematics_MW}


In this work, we have demonstrated that accounting for the deformed dark matter haloes of the Milky Way and LMC is crucial for getting accurate cross section constraints from gamma ray searches. Indeed, Figure~\ref{fig:LMC_impact_on_MW} shows that the change to the constraint from including the deformations (i.e. the difference between the black and red curves) is comparable to the uncertainty on the constraint due to uncertainties in the Milky Way's mass profile (i.e. the grey band). In addition to the effect of accounting for deformations, we also see that precise constraints also depend on how we model the Milky Way and LMC system. The reason for this is that these simulations span a wide range of Milky Way masses and concentrations which affect the strength of the dark matter deformations. In particular, the simulations from \cite{Donaldson+2022} have more massive and concentrated Milky Way haloes which deform less than the simulated Milky Way halo in \cite{2022arXiv220501688L}. For reference, we note that the model in \cite{2022arXiv220501688L} appears similar to models in the literature of the Milky Way-LMC interaction \citep{Garavito-Camargo+2019,Rozier2022}. Given this range of possibilities, we argue that the deformations can be considered as a source of systematic uncertainty on the inferred cross section until they are better characterized.

We note that there are also other physical effects which have altered the Milky Way's dark matter halo and could affect the cross section constraints in similar ways. For example, the Gaia-Sausage/Enceladus \citep[GSE,][]{Belokurov+2018,Helmi+2018} merger likely brought a substantial amount of dark matter into the inner Milky Way which may still not be phase-mixed \citep[e.g.][]{Naidu+2021,Han+2022}. Accounting for this dark matter would likely also lead to changes in the cross section constraints. 

\subsection{Systematic uncertainties affecting the study of the LMC}
\label{sec:discussion_systematics_LMC}



In Sec.~\ref{sec:lat_analysis_LMCroi} we cautioned that the use of the data-driven astrophysical templates for the LMC may artificially drive the DM bounds towards tighter contraints, which we aimed to avoid via the design of the analysis pipeline and the size of the chosen ROI. The \Fermi-LAT collaboration's search for DM in the LMC \citep{Buckley:2015doa} with five years of data mitigated this effect in a different manner. To exemplify the results of this study, we show a selection of upper limits in the right panel of Fig.~\ref{fig:LMC_ROI_study} displayed in three shades of blue representing different choices for the static LMC DM profile. These profiles are tuned to fit the rotation curve of the LMC as traced by stars and gas. The profile dubbed \textsc{nfw-med} is very similar to the initial LMC profiles in the NFW, HERN and Erkal19 simulations on which we base our work. In fact, the cyan line in Fig.~\ref{fig:LMC_ROI_study} is comparable to the results from the dynamical LMC halo (red lines) but also to the bounds from the static DM halo profiles from \cite{Regis:2021glv}. The deviation of the latter bounds can be explained with the enlarged ROI compared to \cite{Buckley:2015doa}. Thus, we find  corroborating evidence that our constructed analysis pipeline is not severely affected by a bias due to the astrophysical LMC templates.

The left panel of Fig.~\ref{fig:LMC_ROI_study} illustrates and quantifies the level of the expected induced variation of upper limits on thermal DM when deformations of the LMC DM halo are included. The implications for the profiles of the obtained upper limits on the annihilation cross section vary between individual simulation suites. Including the LMC halo's deformations may either improve the constraints or weaken them by up to a factor of two. As pointed out in Sec.~\ref{sec:discussion_systematics_MW}, the simulations themselves exhibit an intrinsic uncertainty regarding initial conditions and the definition of the MW/LMC morphology. This uncertainty consequently translates into a range of the \Jf-factor maps of the LMC compatible with the simulated MW-LMC dynamics.

Regarding the obtained upper limits, however, the effect of deviations from spherical symmetry (a defining feature of the monopole of the BFE) must be understood on a case-by-case basis since the DM signal is degenerate with some of the astrophysical background gamma-ray components in the LMC ROI. Deformations of the LMC halo result in asymmetries of the \Jf-factor maps, which may help to break (or worsen) these degeneracies. Considering, for example, the almost spatially uniform isotropic background it is clear that asymmetries in the \Jf-factor maps greatly reduce its degeneracy with the DM template. On the flip-side, it is also conceivable that the morphology of a deformed LMC halo increases existing degeneracies as is the case in the NFW+NFW simulation from \cite{2022arXiv220501688L}.

The degeneracies with the astrophysical background templates also explain the differences between the results for static NFW profiles and simulated LMC haloes displayed in the right panel of Fig.~\ref{fig:LMC_ROI_study}. On one hand, the simulation results exhibit less peaked DM density profiles towards the centre of the LMC than the profiles from \cite{Regis:2021glv}, thus reducing the features that may break degeneracies with background templates. On the other hand, the spectral shape of the DM signal is relevant too in order to improve the constraining power of the analysis. For light DM, for instance, the degeneracy can be broken by the information from the DM gamma-ray annihilation spectrum that shows a cutoff around the DM mass. This cutoff falls within the sweet spot of the LAT's sensitivity. Heavy DM with $m_{\chi} > 100\unit{GeV}$, in contrast, features a break in the spectrum at tens of GeV where the LAT sensitivity starts to decrease. Hence, in this particular case, the spectral shape of the annihilation signal does not contribute as much to breaking the degeneracy with the background.
\section{Conclusions}
\label{sec:conclusions}
In the present work, we used state-of-the-art simulations of the
MW encounters with the LMC to assess how the deformations of the MW and LMC dark matter haloes affect indirect detection for DM.

First, we focused on high Galactic latitudes and performed a search for a DM annihilation signal
in twelve years of \Fermi-LAT data.
Since no significant signal was found (regardless of the DM spatial distribution adopted), we set 95\% C.L.~constraints
on the DM annihilation cross section. 
In particular, we modelled the DM distribution in the Galactic halo following a recent MW mass modelling~\citep{2017MNRAS.465...76M}.
The mass modelling of the MW comes with non-negligible uncertainties. We propagated this
uncertainty on our final limits, and provided an uncertainty band which reflects it. 
Moreover, we verified that the systematic uncertainties from IE modelling are indeed mild as they induce a variance of the derived bounds by at most a factor of two for DM masses at the light and heavy end of the probed mass range. The optimal ROI sizes for each of the employed IE models are largely overlapping rendering a direct comparison sensible (see Appendix~\ref{app:iem_limits} for more details).
The limits in Fig.~\ref{fig:McMillan_limits} represent
the most up-to-date and robust limits on DM at high latitudes derived with \Fermi-LAT data.
Our high-latitude DM limits are stronger than the results obtained by \cite{Zechlin:2017uzo} from their most constraining energy band from 1 to 2 GeV (c.f.~orange line in the left panel of Fig.~\ref{fig:McMillan_limits}). As noted in Sec.~\ref{sec:ULIM_McMillan2017}, the authors performed their analysis in an ROI overlapping with ours but of reduced size. Thus, our enlarged ROI improves the constraining power allowing us to exclude thermal DM for masses $m_{\chi} \lesssim 40$ GeV, while \cite{Chang:2018bpt} report a slighter stronger bound for DM below 200 GeV and an exclusion for $m_{\chi} \lesssim 70$ GeV. To this end, the authors have derived an ROI that is rather disjoint with ours. It is closer to the Galactic centre and it takes into account gamma-ray data from the position of the FBs. The latter two differences can easily explain the increased constraining power compared to our study. We notice that our bounds do not account
for the presence of sub-halos, which are expected to boost the annihilation signal, and therefore
strengthening the limits, especially if very highly concentrated, see e.g.~\cite{2022arXiv220705082D}.

In the right panel of Fig.~\ref{fig:McMillan_limits}, we place our MW outer halo bounds in the context of existing constraints on the parameter space of thermal DM derived from different targets and cosmic-ray channels. The light purple line indicates gamma-ray constraints from the observation of dwarf spheroidal galaxies with space-borne and ground-based instruments. The results of this joint analysis \citep{Hess:2021cdp} are the most state-of-the-art constraints from dwarf spheroidal galaxies using traditional inference techniques. While this set of exclusion limits outperforms our bounds over the entire probed mass range, they are affected by non-negligible systematic uncertainties due to both modelling of the DM distribution in these systems and the background modelling in and around dwarf spheroidal galaxies. As for the latter, the authors of \cite{Calore:2018sdx,Alvarez:2020cmw} designed an analysis that incorporates background modelling systematic effects, which weaken the upper limits on the DM annihilation cross section of about a factor of three. These limits (based only on classical dwarf spheroidal galaxies) are shown as a purple line. In this case, constraints from the outer MW halo are stronger than the gamma-ray limits from dwarf spheroidal galaxies. The variance induced by the uncertainty of the outer MW halo profile is hence less pronounced than the impact of background modelling uncertainties in dwarf spheroidal galaxy studies.
We stress that we assumed a very conservative approach in the selection of the
region of interest for the analysis by requiring a strict compatibility between the statistical expectations derived from Poisson realisations of the baseline fit and the true \Fermi-LAT data. This method yields reliable and robust limits but reduces the potential constraining power of the full data set and limits the accessible dynamically generated features of the MW-LMC system. As indicated by the right panel of Fig.~\ref{fig:jfactor_comparison_LMC}, the northern sky may be indeed a more promising target to explore the signatures of the MW-LMC interaction. In this part of the sky, the LMC is expected to provoke a response in the forward direction of its orbit. However, finding a good agreement between astrophysical model and 
data is more challenging and needs better theoretical refinements. For example, if we relax the strict constraint of considering only data at high latitudes, we find that adding to the optimal ROI in Sec.~\ref{sec:upperlimits} a counterpart in the northern hemisphere defined by $-102^{\circ} < \ell < 102^{\circ}$ and $b > 16^{\circ}$ (excluding the FBs region) yields the best accordance -- although being far from perfect -- between the expectations derived from the baseline fit and the true \Fermi-LAT data. The improvement we can achieve in this way is as large as a factor of 3 for either a static or a simulated MW halo profile (see details in Appendix~\ref{app:newROI_limits}).
In contrast, constraints (green band) from radio searches towards the LMC as found in \cite{Regis:2021glv} show a larger intrinsic uncertainty than ours. Even in a conservative scenario, however, the bounds on thermal DM from the LMC's emission in radio light are stronger over the considered mass range. It should be noted that these results are derived assuming a static LMC DM profile while dynamical effects -- as we have studied in this work -- may alter the picture.
The authors of \cite{DiMauro:2021qcf} have derived a set of upper limits (dark blue) comparable to the radio LMC bounds considering the latest AMS-02 antiproton data release. 
Anti-proton constraints seem robust and only mildly affected by modelling uncertainties (DM halo profile, comsic-ray propagation). In the future, these bounds can be further strengthened with a re-calibrated
prediction of secondary cosmic rays to achieve a better agreement with new data, which is further discussed in~\cite{Calore:2022stf}.

Thanks to the uncertainty band from the MW gravitational potential modelling,
we quantified the significance of the MW-LMC dynamics' impact on indirect DM detection.
For the set of simulations adopted in this work, we found that the MW-LMC interaction 
does not strongly affect the upper limits on thermal DM. The obtained variations are within
a factor of 1.3, which we find in the Erkal19 simulation that generally yields the most pronounced dynamical responses.

%

The MW-LMC interactions also affect the mass distribution in the LMC itself -- and the ensuing DM annihilation signal.
Therefore, we derived bounds on DM annihilation from the region around the LMC as well, with dedicated 
modelling of the astrophysical backgrounds. 
The limits derived from the LMC ROI are largely consistent with the previous bounds derived by the \Fermi-LAT collaboration for the case of a static NFW profile as discussed in Sec.~\ref{sec:discussion_systematics_LMC}. The level of systematic uncertainty of the constraints on thermal DM caused by the dynamical deformation of the LMC halo is not of the same order of magnitude as the one caused by the allowed range of static DM halo profiles that can reproduced the stellar rotation curve of the LMC as illustrated by the range of the blue-shaded lines in Fig.~\ref{fig:LMC_ROI_study}. However, we note that the initial states of the LMC in all simulations utilised in this work did not aim at bracketing the full margin of inner halo profiles consistent with stellar data. We can thus not quantify how the LMC's internal structure is affected by its passage through the MW in case of an aggressive assumption like the \textsc{sim-med} parametrization in \cite{Buckley:2015doa}. Refined simulations with a wider range of inner DM halo slopes may be warranted to assess the full implications of the MW-LMC dynamics for indirect searches towards the LMC. Eventually, the effect may even alter the prospects of radio searches in the central region of the LMC which consequently relax or tighten the already stringent bounds on thermal DM reported by \cite{Regis:2021glv}.

In conclusions, we have shown that high Galactic latitudes have the potential to  be the leading target for DM searches in gamma rays in the future.
A crucial step in this direction will be the optimisation of interstellar emission 
models thanks to, for example, machine learning techniques and verification schemes, see e.g.~\cite{Storm:2017arh,Mishra-Sharma:2021oxe, Caron:2021wmq} in the context of background model optimisation in the inner Galaxy.
If we focus on the LMC region, a leap forward in the understanding
of the astrophysical emission (and therefore a better modelling of the LMC astrophysical templates) is expected thanks to the up-coming observations
of the Cherenkov Telescope Array, CTA, of this particular region as outlined in \cite{CTAConsortium:2017dvg}.

Looking forward, future work with stellar streams and other tracers will allow the community to robustly measure the dark matter deformations of the Milky Way and LMC dark matter haloes through their gravitational effects. Indeed, \cite{Shipp+2021} show that the streams have their closest approaches with the LMC at different times, giving hope to the idea that the time dependence can be measured. Once these deformations are measured, they can be folded into the analysis, as we have done in this work, to derive the most accurate annihilation cross sections. On the flip side, if annihilating dark matter is detected in gamma rays then we will be able to directly measure the deforming dark matter haloes of the Milky Way and LMC.

\section*{Acknowledgements}
\addcontentsline{toc}{section}{Acknowledgements}
The work of C.E. is supported by the ``Agence Nationale de la Recherche'', grant n. ANR-19-CE31-0005-01 (PI: F. Calore). M.S.P is partially supported by grant Segal ANR-19-CE31-0017 of the French Agence Nationale de la Recherche (https://secular-evolution.org).

For the purpose of open access, the author has applied a Creative Commons Attribution (CC BY) licence to any Author Accepted Manuscript version arising
from this submission.\\
\emph{Software:} \textsc{astropy} \citep{2013A&A...558A..33A, 2018AJ....156..123A,TheAstropyCollaboration2022}, \textsc{clumpy} (version 3) \cite{charbonnier2012clumpy, bonnivard2016clumpy, 2019CoPhC.235..336H}, \textsc{fermi science tools} \citep{2019ascl.soft05011F}, \textsc{healpy} \citep{2005ApJ...622..759G, Zonca2019}, \textsc{iminuit} \citep{iminuit}, \textsc{jupyter} \citep{soton403913}, \textsc{matplotlib} \citep{4160265}, \textsc{numpy} \citep{2020Natur.585..357H}, \textsc{scipy} \citep{2020NatMe..17..261V}
\section*{Data Availability}
\addcontentsline{toc}{section}{Data Availability} A python interface to integrate orbits
and access the expansion model for the Erkal19 simulation can be found here:
\url{https://github.com/sophialilleengen/mwlmc}. The code to reproduce the results of the \Fermi-LAT gamma-ray analysis is provided at:
\url{https://github.com/ceckner/MWLMCGammaRays}.

\bibliographystyle{mnras}
\bibliography{MWLMC.bib} 


\clearpage

\appendix

\section{Astrophysical fore- and back-ground components}
\label{app:astro_comp}
We here provide more details about the astrophysical fore- and back-ground components used to fit the gamma-ray sky.

\begin{itemize}
    \item \textbf{interstellar emission (IE)}: A diffuse Galactic gamma-ray source emerging due to very-high energy, charged cosmic rays impinging on particles of the MW's interstellar medium, dust and radiation fields. The dominant processes that create gamma rays are $\pi^0$-decay, Bremsstrahlung and IC scattering. The modelling of this contribution is subject to many uncertainties so that we include five different models that aim to quantitatively characterise the IE. Two IE models (henceforth called \emph{Lorimer I} and \emph{Lorimer II}) are taken from the set of realisations considered in the ``1st Fermi LAT Supernova Remnant Catalog''\footnote{The model files have been made public by the \Fermi-LAT collaboration at: \url{https://fermi.gsfc.nasa.gov/ssc/data/access/lat/1st_SNR_catalog/}.} \cite{Acero:2015prw}. While a detailed description of the models' preparation and their respective properties are given in the cited catalogue publication, we restrict ourselves to a brief summary of their main characteristics relevant to this work: The distribution of primary cosmic rays is linked to the distribution of pulsars as analysed and discussed in \cite{Lorimer:2006qs}. They are confined in a volume with propagation height $z=10$ kpc whereas the spin temperature of the interstellar medium is assumed to be $T_s = 150/1\cdot10^5$ K (Lorimer I/II). The interstellar medium's gas content is split in atomic hydrogen \ce{H} and \ce{CO} maps. The latter serves as a proxy for the distribution of molecular hydrogen \ce{H2}. The total volume of the gas maps is decomposed into four Galactocentric annuli with respective extensions: 0-4 kpc: ``ring 1'', 4-8 kpc: ``ring 2'', 8-10 kpc: ``ring 3'' and 10-30 kpc: ``ring 4''. Since the IE from different rings contributes to different parts of the gamma-ray sky, we make use of this decomposition in our fitting strategy outlined in Sec.~\ref{sec:fitting_opti}. We note that the properties of IE model Lorimer I are comparable to those of the official diffuse background model of the \Fermi-LAT derived in connection with the 4FGL catalogue \cite{Fermi-LAT:2019yla,2020arXiv200511208B}.Therefore, we declare Lorimer I our benchmark IE model.\linebreak
    The three remaining models -- called \emph{foreground model A, B} and \emph{C} -- are adopted\footnote{All model files are stored in the \Fermi-LAT collaboration's public data archive: \url{https://www-glast.stanford.edu/pub_data/845/}.} from a careful and detailed study of the diffuse extragalactic gamma-ray background conducted by the \Fermi-LAT collaboration \cite{Ackermann:2014usa}. In contrast to Lorimer I and II, these IE instances exhibit the advantage to be directly prepared for the study of a large-scale emission component at high Galactic latitudes -- a feature aligned with the aim of our analysis. Again, we refer to the cited publication to learn more about the exact composition of the models.
    \item \textbf{isotropic diffuse gamma-ray background (IGRB)}: This large-scale contribution to the gamma-ray sky is spatially isotropic, hence following the spatial structure of the LAT's exposure, and generated by the collective emission of distant extragalactic gamma-ray emitters too faint to be resolved individually. We adopt the IGRB spectrum associated with the selected LAT data's event class and type (see Sec.~\ref{sec:lat_data_selection}) provided by the Fermi Science Tools\footnote{The IGRB spectrum files are also provided at \url{https://fermi.gsfc.nasa.gov/ssc/data/access/lat/BackgroundModels.html}}. We stress that these spectra are only valid in combination with the official diffuse background model of the \Fermi-LAT collaboration. Since our fit model (Eq.~\ref{eq:model_eq}) renormalises each background component per energy bin, this restriction is, however, irrelevant for our cause. 
    \item \textbf{resolved point-like and extended gamma-ray sources}: Based on a 10-year data set, the \Fermi-LAT collaboration has published an extensive catalogue, 4FGL-DR2, \citep{Fermi-LAT:2019yla,2020arXiv200511208B} of all resolved and localised point-like or extended gamma-ray emitters in or outside of the MW. We include all of these sources with their respective properties reported in the catalogue in our analysis. A detailed description of the treatment of these sources in each analysis step is given in the following Sec.~\ref{sec:fitting_opti}.
    \item \textbf{Fermi Bubbles (FB)}: An extended, hourglass-shaped diffuse gamma-ray source, which is present above and below the Galactic disc up to high Galactic latitudes. We adopt the spatial morphology of the FBs as derived in \cite{TheFermi-LAT:2017vmf} whereas its spectrum is taken to be a log-parabola $\frac{\mathrm{d}N}{\mathrm{d}E}=F_{0}\left(\frac{E}{E_{0}}\right)^{-\alpha-\beta\ln\!{\left(E/E_{0}\right)}}$ described by the parameters $F_{0}=5\times10^{-10}\;\mathrm{ph}\,\mathrm{cm}^{-2}\,\mathrm{s}^{-1}\,\mathrm{MeV}^{-1}$, $\alpha$ = 1.6, $\beta$ = 0.09 and $E_{0}$ = 1 GeV as reported in \cite{Herold:2019pei}.
    \item \textbf{LoopI}: A large-scale, loop-like structure exhibiting a diffuse gamma-ray emission mainly concentrated in the northern hemisphere above the Galactic disc. We include in our model the spatial and spectral characterisation of LoopI given in \cite{Wolleben:2007pq}. This particular model also features a non-vanishing contribution in the southern hemisphere of the gamma-ray sky.
    \item \textbf{gamma-ray emission induced by the Sun and Moon}: Since the LAT is in constant observation mode, both the Sun and the Moon cross the field of view of the telescope. Along their orbital trajectory they contribute a non-negligible gamma-ray emission. We make use of dedicated routines within the Fermi Science Tools\footnote{A technical description of the software developed for this task is given at: \url{https://fermi.gsfc.nasa.gov/ssc/data/analysis/scitools/solar_template.html}} to derive a flux model for the gamma-ray emission of these two celestial bodies tailored towards the selected LAT data (see \cite{2013ICRC...33.3106J}).
\end{itemize}

\section{Selected bright 4FGL-DR2 sources}
\label{app:bright_4FGL}

We report in Tab.~\ref{tab:bright_4fgl} all sources in 4FGL-DR2 whose energy flux (100 MeV - 100 GeV) is above the threshold of $4\cdot10^{-10}\;\left[\mathrm{MeV}\,\mathrm{cm}^{-2}\,\mathrm{s}^{-1}\right]$. These sources are considered as individual templates during the iterative fit (see Sec.~\ref{sec:fitting_opti}) designed to derive a baseline model of the gamma-ray sky solely comprised of the astrophysical background components introduced in Sec.~\ref{sec:bkg_selection}. We state their name in 4FGL-DR2, their position in Galactic coordinates, the nominal energy flux and the source class using abbreviations from \cite{Fermi-LAT:2019yla}.

\begin{table*}
    \centering
    \begin{tabular}{l l c c}
         \hline
         4FGL source name & $\left(\ell, b\right)\;\left[^{\circ}\right]$ & $E_{100}\;\left[\mathrm{MeV}\,\mathrm{cm}^{-2}\,\mathrm{s}^{-1}\right]$ & source class\\
         \hline
         \hline
4FGL J0835.3-4510 & $\left(263.6, -2.8\right)$ & $9.4\cdot10^{-9}$ & PSR\\
4FGL J0633.9+1746 & $\left(195.1, 4.3\right)$ & $4.2\cdot10^{-9}$ & PSR\\
4FGL J0534.5+2200 & $\left(184.6, -5.8\right)$ & $1.4\cdot10^{-9}$ & PSR\\
4FGL J1709.7-4429 & $\left(343.1, -2.7\right)$ & $1.4\cdot10^{-9}$ & PSR\\
4FGL J2028.6+4110e & $\left(79.6, 1.4\right)$ & $1.1\cdot10^{-9}$ & SFR\\
4FGL J2253.9+1609 & $\left(86.1, -38.2\right)$ & $1.0\cdot10^{-9}$ & FSRQ\\
4FGL J2021.5+4026 & $\left(78.2, 2.1\right)$ & $8.4\cdot10^{-10}$ & PSR\\
4FGL J1836.2+5925 & $\left(88.9, 25.0\right)$ & $6.2\cdot10^{-10}$ & PSR\\
4FGL J2021.1+3651 & $\left(75.2, 0.1\right)$ & $5.3\cdot10^{-10}$ & PSR\\
4FGL J2232.6+1143 & $\left(77.4, -38.6\right)$ & $4.9\cdot10^{-10}$ & FSRQ\\
4FGL J1855.9+0121e & $\left(34.7, -0.4\right)$ & $4.9\cdot10^{-10}$ & SNR\\
4FGL J0240.5+6113 & $\left(135.7, 1.1\right)$ & $4.7\cdot10^{-10}$ & HMB\\
4FGL J1256.1-0547 & $\left(305.1, 57.1\right)$ & $4.5\cdot10^{-10}$ & FSRQ\\
4FGL J0617.2+2234e & $\left(189.0, 3.0\right)$ & $4.5\cdot10^{-10}$ & SNR\\
4FGL J1809.8-2332 & $\left(7.4, -2.0\right)$ & $4.4\cdot10^{-10}$ & PSR\\
4FGL J0007.0+7303 & $\left(119.7, 10.5\right)$ & $4.3\cdot10^{-10}$ & PSR\\
4FGL J1104.4+3812 & $\left(179.8, 65.0\right)$ & $4.2\cdot10^{-10}$ & BLL\\
4FGL J1745.6-2859 & $\left(359.95, -0.04\right)$ & $4.2\cdot10^{-10}$ & spp\\
4FGL J1512.8-0906 & $\left(351.2, 40.1\right)$ & $4.2\cdot10^{-10}$ & FSRQ\\
4FGL J0534.5+2201i & $\left(184.6, -5.8\right)$ & $4.1\cdot10^{-10}$ & PWN\\
         \hline
    \end{tabular}
    \caption{Summary table listing bright 4FGL-DR2 sources with an energy flux $E_{100} \geq4\cdot10^{-10}\;\left[\mathrm{MeV}\,\mathrm{cm}^{-2}\,\mathrm{s}^{-1}\right]$ that are fit individually during the iterative fit procedure aimed to derive a baseline model of the gamma-ray sky. The table states the source's name in 4FGL-DR2, its position in Galactic longitude $\ell$ and latitude $b$ in degree, the nominal energy flux $E_{100}$ and the source class using abbreviations from \protect\cite{Fermi-LAT:2019yla}.\label{tab:bright_4fgl}}
\end{table*}

\section{Injected dark matter signal recovery}
\label{app:signal_recovery}

In order to test the performance of the analysis pipeline outlined in Sec.~\ref{sec:fitting_opti} and to verify its trustworthiness, we conduct a simple signal recovery exercise: Utilising the baseline fit with IE model Lorimer I on the all-sky gamma-ray data, we inject a DM signal into the model and try to detect it with the chosen inference method, namely the log-likelihood ratio test statistic. The test statistic in Eq.~\ref{eq:TS_stat_reach} is not suited for this task since it is designed to constrain an alternative hypothesis when the data seems to prefer the background-only hypothesis. Thus, we modify the test statistic for the case of determining the detection significance according to \citep{Cowan:2010js}:

\newpage 

\begin{widetext}
\begin{equation} 
\label{eq:TS_discovery} 
\qquad\qquad\qquad\qquad\qquad\qquad\textrm{TS}_{\mathrm{discovery}}= \begin{cases} -2\min_{\{N_i^{B_j}\}}\left(\ln\!\left[\frac{\mathcal{L}_w\!\left(\left.\bm{\mu}(N^{\mathrm{DM}} = 0,N_i^{B_j}) \right|\bm{n}\right)}{\mathcal{L}_w\!\left(\left.\bm{\hat{\mu}}\right|\bm{n}\right)}\right]\right)\, & \hat N^{\mathrm{DM}} \geq 0\\ 0 & \hat N^{\mathrm{DM}} < 0\rm{,} 
\end{cases} 
\end{equation}
\end{widetext}
where we keep the notation established in Sec.~\ref{sec:stat_framework}. Succinctly put, this test statistic assumes that the preferred hypothesis contains a positive signal with a positive best-fit normalisation $\hat N^{\mathrm{DM}}$. The alternative hypothesis, in contrast, becomes the background-only hypothesis, that is, finding the best-fit background parameters under the assumption that $N^{\mathrm{DM}} = 0$. The likelihood ratio thus quantifies the significance of the detected signal and for a given threshold one may claim a detection at the $\cdot\sigma$ level. We have applied this prescription to ascertain that no significant signal is present in the selected LAT data for all scenarios provided in the main text. Hence, deriving upper limits on the WIMP DM pair-annihilation cross section  is justified. 

For the purpose of our pipeline check, we conduct the following adapted approach:
\begin{itemize}
    \item For the monopole and full BFE MW + LMC haloes of the Erkal19 simulation, we prepare a signal template featuring a DM particle with mass $m_{\chi} = 100\unit{GeV}$ annihilating into $b\bar{b}$ pairs. 
    \item For discrete points in the range of annihilation cross section  $N^{\mathrm{DM}} = \langle\sigma v \rangle \in \left[10^{-28}, 10^{-23}\right]\unit{cm}^3\unit{s}^{-1}$, we inject the signal template with normalisation $N^{\mathrm{DM}} $ into the baseline fit of Lorimer I.
    \item Drawing 200 Poisson realisations of the baseline fit + signal gamma-ray sky, we perform a maximum likelihood fit with respect to this mock data and save the retrieved best-fit signal strength $\hat N^{\mathrm{DM}}$.
    \item Again, using 200 Poisson realisations of the baseline fit + signal gamma-ray sky, we derive the upper limit on the DM annihilation cross section  that we may set given the presence of a signal in the utilised mock data.
\end{itemize}

\begin{figure*}
\centering
\includegraphics[width=0.7\textwidth]{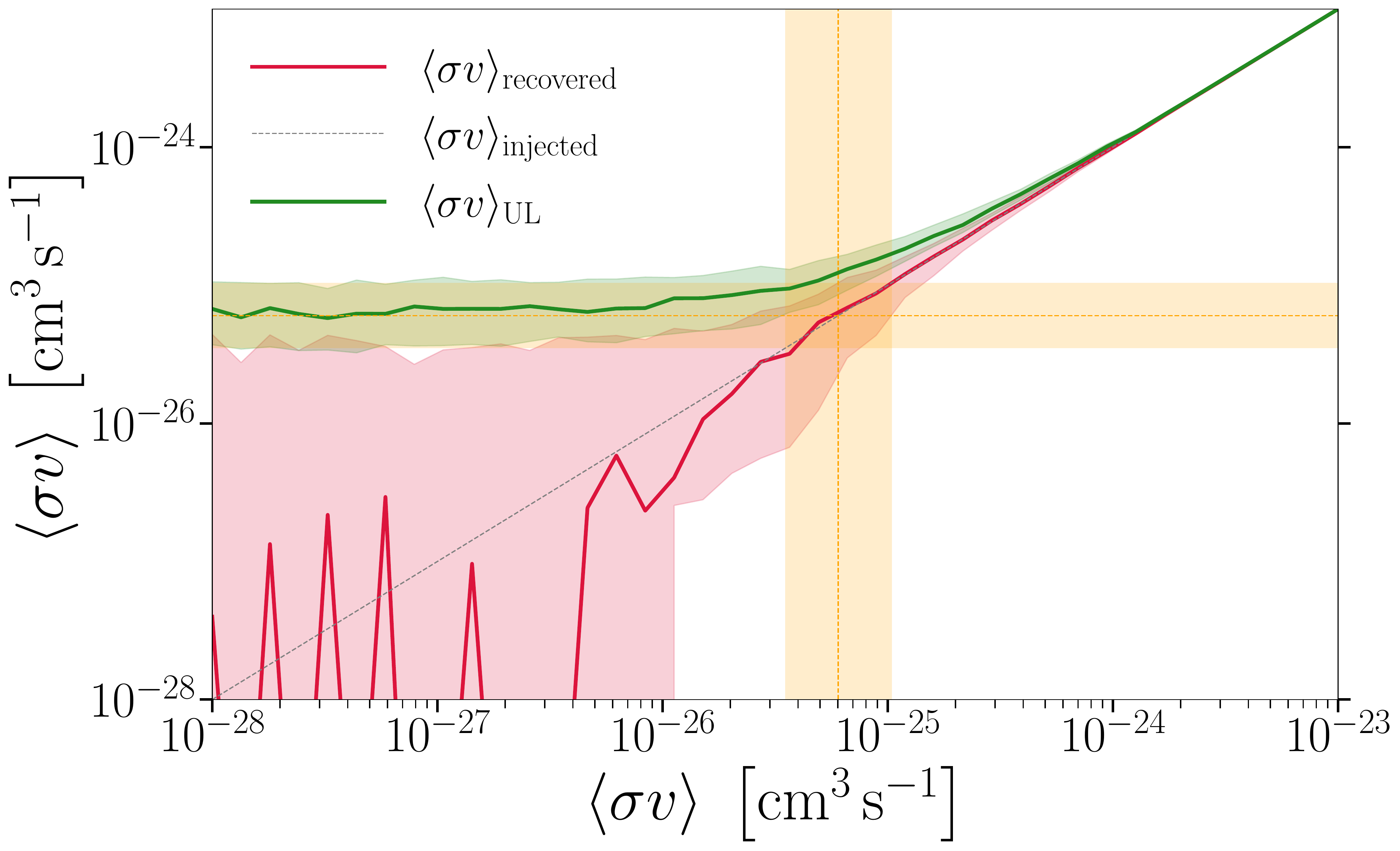}
\caption{Signal recovery test for WIMP particle DM of mass $m_{\chi} = 100\unit{GeV}$ annihilating into $b\bar{b}$ pairs and a DM halo featuring the full BFE of the Erkal19 simulation to model the spatial morphology. The signal template has been injected with a normalisation $N^{\mathrm{DM}}$ following the dashed black line into the baseline fit for IE model Lorimer I. The solid red line denotes the median value of the recovered best-fit signal normalisation for 200 Poisson realisations drawn from the mock data while the red band illustrates the expected scatter ($68\%$ containment) of the DM best-fit parameter. The solid green line represents the median $95\%$ C.L.~upper limit that follows from the same mock data with injected DM template; the $68\%$ containment is shown as green band. Finally, the solid orange line is the median $95\%$ C.L.~upper limit on mock data that does not contain a signal contribution. Again, the orange band denotes the associated $1\sigma$ scatter. \label{fig:signal_recovery_100GeV}}
\end{figure*}

The results of this sanity check are displayed in Fig.~\ref{fig:signal_recovery_100GeV}, which shows the median/scatter of the recovered signal normalisation (red solid/shaded band) and the median/scatter of the upper limits (green solid/shaded band) as a function of the injected signal's strength. To guide the eye, we also denote the upper limit on this particular particle DM model derived from the baseline fit without injected signal as orange line; its $68\%$ containment band is given as an orange-shaded band. 

The general expectation for this kind of pipeline check is to recover the injected signal strength; the higher the signal strength the higher the confidence of recovering the signal, i.e.~the smaller the observed scatter of the best-fit parameter. Moreover, for extremely small annihilation cross sections the upper limits with respect to mock data that contain the signal should asymptotically approach the corresponding upper limits with respect to the baseline fit standalone. Both of these features we can confirm with our sanity check. In addition, we find that the solid red line approaches the dashed black line for cross section  values below the nominal upper limit marked by the vertical orange line. This suggests that the pipeline is capable of theoretically detecting a signal less luminous than the obtained upper limit. At the same time, the solid green line starts to deviate from the horizontal orange line when the injected signal strength can be recovered. We can hence conclude that the constructed analysis pipeline works as intended and it is suitable to perform the envisaged tasks.


\section{Dark matter constraints for different interstellar emission models}
\label{app:iem_limits}

\begin{figure*}
\centering
\includegraphics[width=0.49\textwidth]{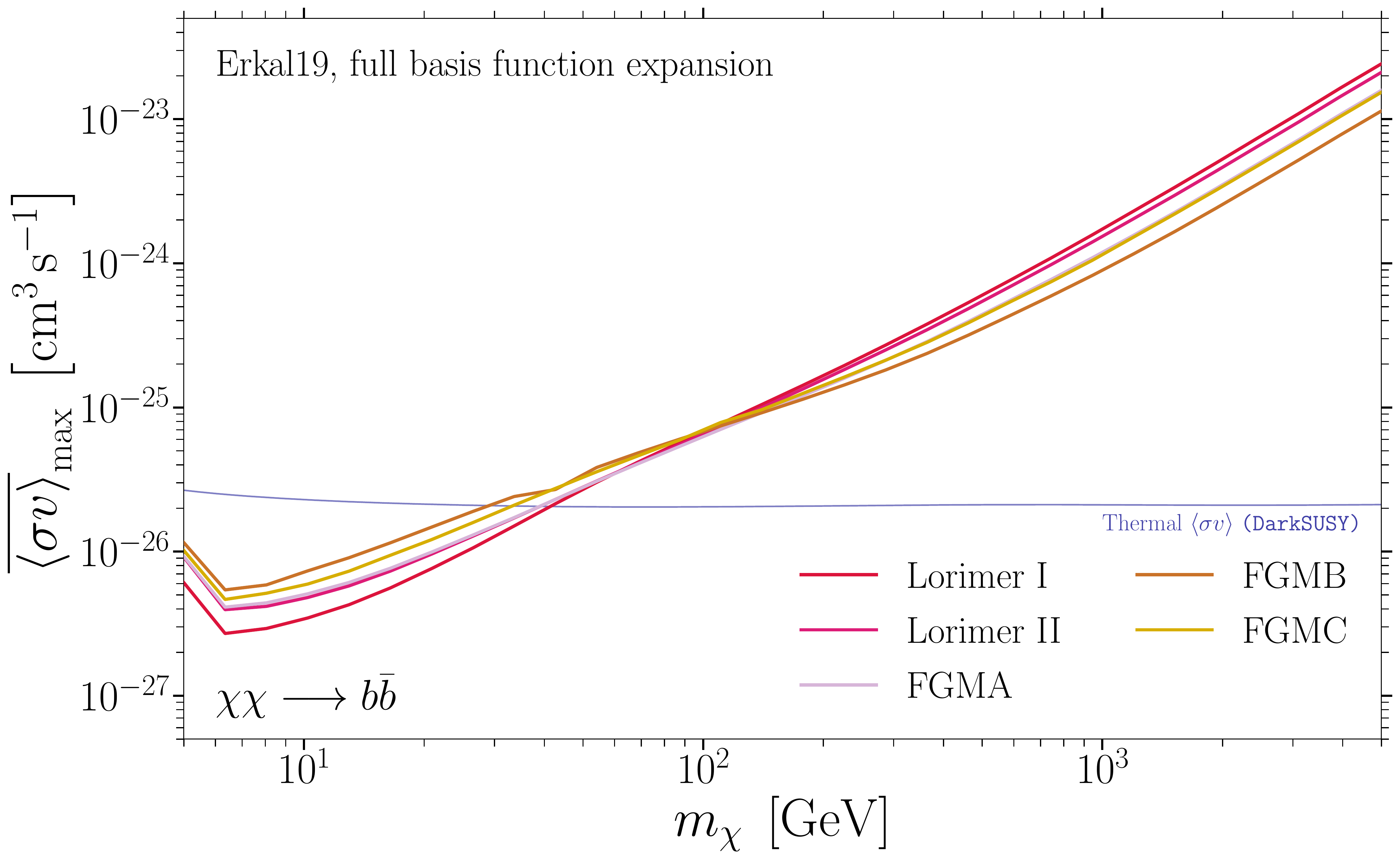}
\hfill
\includegraphics[width=0.49\textwidth]{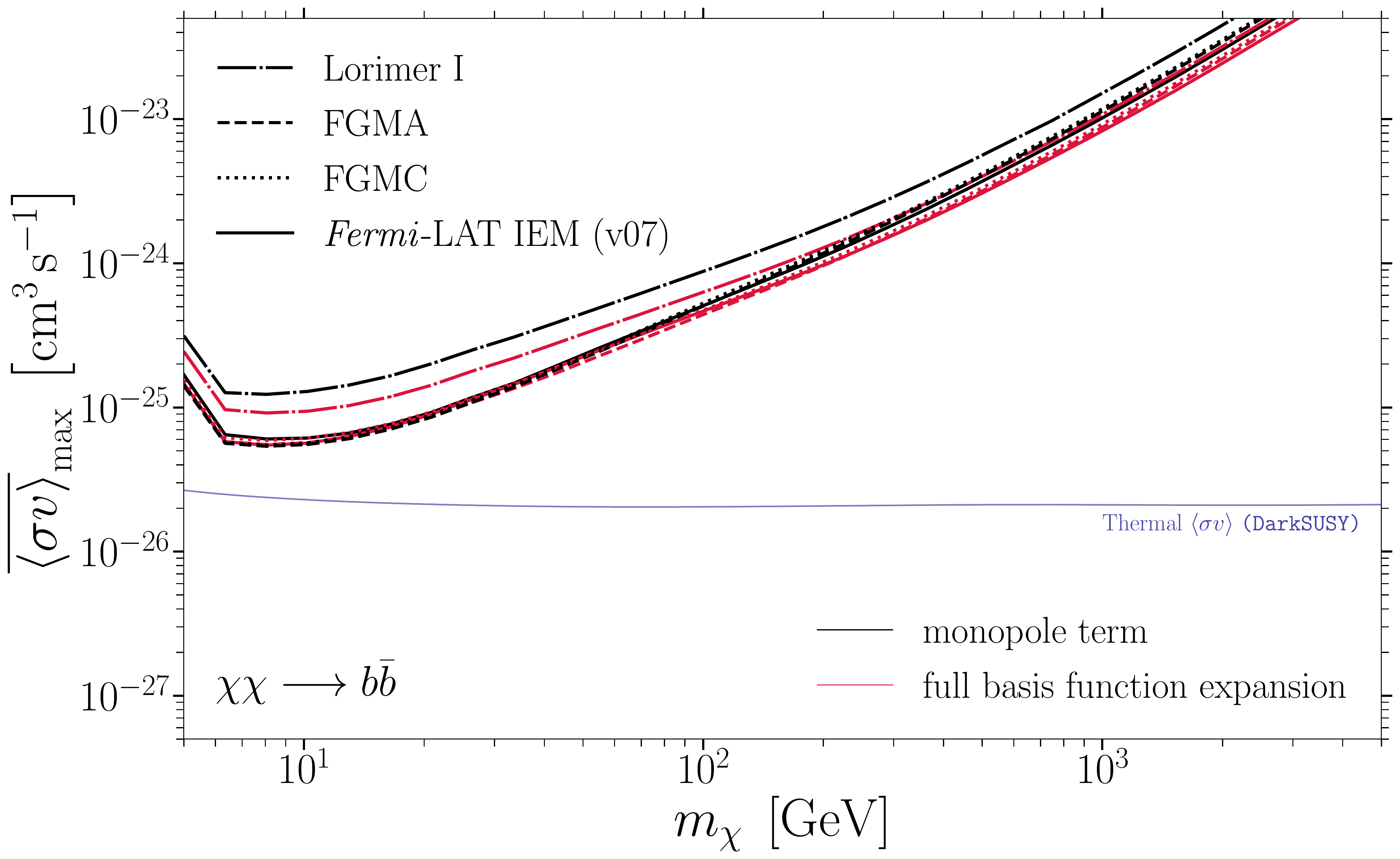}
\caption{$95\%$ C.L.~upper limits on the velocity-weighted thermally averaged DM pair-annihilation cross section  depending on the assumed DM mass $m_{\chi}$ for the prompt gamma-ray emission from DM annihilating into $b\bar{b}$ final states. (\emph{Left}: \textbf{analysis of the outer MW halo}) Variance of the observed upper limits due to changing the IE model. As an illustrative example, we display the signal morphology following the \Jf-factor of the combined MW and LMC halo according to the Erkal19 simulation including the full BFE. Note that each IE model requires a different optimised ROI. The corresponding values are provided in Tab.~\ref{tab:iem_rois}. (\emph{Right}: \textbf{analysis of the LMC region}) Impact of varying the IE model used to derive the constraints on DM. We show four distinct IE models: Lorimer I (dashed), \Fermi-LAT IE model (v07) (solid), FGMA (dashed) and FGMC (dotted). Black lines represent the unperturbed LMC halo whereas red lines display the respective scenario including dynamical effects based on the halo profiles derived from the simulation based on a Hernquist profile for the MW and an NFW profile for the LMC as initial conditions from Ref.~\protect\cite{Donaldson+2022}. The optimal ROI sizes for each scenario are stated in Tab.~\ref{tab:lmc_rois_iem}. \label{fig:iem_uncertainty_main}}
\end{figure*}

The main results in Sec.~\ref{sec:upperlimits} of our analysis of the MW-LMC system rely on the choice of a particular benchmark IE model. However, it is well known that the characterisation of the interstellar emission is not perfect and a great deal of data- and observation-driven as well as simulated models have been put forward to achieve a sufficiently good agreement between reality and theoretical understanding plus observational data. On one hand, since our study mostly focuses on the high-latitude in the southern hemisphere of the gamma-ray sky, our results are less severely affected by the IE directly related to the emission originating from the interstellar medium along the Galactic disc. On the other hand, gamma-ray emission due to IC scattering events is particularly hard to model and exhibits large uncertainties so that the impact of the IE model on our results is certainly non-negligible. As motivated in Sec.~\ref{sec:bkg_selection}, we aim to assess the systematic uncertainty of our results due to the modelling of the IE via alternative models that supplement the benchmark choices for either the outer MW halo study or the dedicated study of the LMC surroundings. In Fig.~\ref{fig:iem_uncertainty_main}, we provide a set of $95\%$ C.L.~upper limits on the DM pair-annihilation cross section  (again, with respect to the channel $\chi\chi\rightarrow b\bar{b}$) for different choices of the IE model. 

The left panel of this figure concerns the impact of the IE on the study of the outer MW halo, where we exemplify the performance of all five IE models with the signal morphology according to the \Jf-factor map of the Erkal19 simulation including all terms of the BFE for both DM haloes. We checked that the choice of the simulation is not important to quantify the impact of the IE uncertainty; the other four simulations as well as the static MW halo from \cite{2017MNRAS.465...76M} yield very similar results. Using a different IE model than Lorimer I as in the main text, requires us to re-perform the ROI optimisation, which yields optimal ROI sizes detailed in Tab.~\ref{tab:iem_rois} for the respective case. As concerns the comparison of upper limits derived from real data and the baseline fit, we find consistency for almost the entire probed DM mass range at the $2\sigma$ level and even better for FGMA to FGMC; an expected behaviour since these three models were initially created to facilitate searches at high latitudes. A slight fluctuation to stronger constraints for light DM with masses below $\lesssim 20\unit{GeV}$ is a common feature among all probed IE models. Overall, the numerical values of the derived constraints are comparable to the corresponding upper limits for the same simulation set in Fig.~\ref{fig:LMC_impact_on_MW} but IE model Lorimer I. The apparent differences certainly partially arise because of the varying optimal ROI sizes that reduce or increase the total \Jf-factor.

The right panel of Fig.~\ref{fig:iem_uncertainty_main} addresses the question of the IE model's impact on the study of the LMC region. As explained in Sec.~\ref{sec:lat_analysis_LMCroi}, we include for this particular task the Galactic diffuse background model of the \Fermi-LAT collaboration. Besides, we do not show the upper limits for FGMB, as it appears to behave exactly like FGMC, and Lorimer II, which does not provide a good fit to the data in general. In this case, we use the simulation based on a Hernquist profile for the MW and an NFW profile for the LMC as initial conditions from Ref.~\cite{Donaldson+2022}. This example suffices to quantify the impact of the IE modelling as the remaining LMC halo models exhibit similar behaviour. Although we display the set of upper limits for each particular IE model on the same plot, the underlying optimised ROI sizes are different and all applied values are given in Tab.~\ref{tab:lmc_rois_iem}. However, the differences in those sizes are not noticeably altering the total \Jf-factor we are probing so that a common plot is justified. As it turns out, varying the IE has a remarkably small effect on the resulting constraints on WIMP DM. The only exception being the IE model Lorimer I, which induces a deterioration of the limits by a factor of $\sim2$ compared to the other three IE model instances. Hence, the results of the dedicated LMC study are robust against variations in the IE modelling. 


\begin{table*}
\begin{centering}
\begin{tabular}{c c c}
\hline 
\multirow{3}{*}{IE model} & only monopole & multipole expansion\tabularnewline
\cline{2-3} \cline{3-3} 
 & ROI size & ROI size\tabularnewline
 & \Jf-factor {[}GeV$^2$cm$^{-5}${]} & \Jf-factor {[}GeV$^2$cm$^{-5}${]}\tabularnewline
\hline 
\hline
\multirow{2}{*}{Lorimer I} & $27.2^{\circ}\times 27.2^{\circ}$  & $27.2^{\circ}\times27.2^{\circ}$ \tabularnewline
 & $1.38\times10^{20}$ & $1.62\times10^{20}$\tabularnewline
\hline 
\multirow{2}{*}{\Fermi-LAT IE model (v07) } & $28.4^{\circ}\times28.4^{\circ}$ & $29.0^{\circ}\times29.0^{\circ}$\tabularnewline
 & $1.42\times10^{20}$ & $1.68\times10^{20}$\tabularnewline
\hline 
\multirow{2}{*}{FGMA} & $29.5^{\circ}\times29.5^{\circ}$ & $29.5^{\circ}\times29.5^{\circ}$\tabularnewline
 & $1.45\times10^{20}$ & $1.70\times10^{20}$\tabularnewline
\hline 
\multirow{2}{*}{FGMC} & $29.4^{\circ}\times29.4^{\circ}$ & $29.4^{\circ}\times29.4^{\circ}$\tabularnewline
 & $1.45\times10^{20}$ & $1.79\times10^{20}$\tabularnewline
\hline 
\end{tabular}
\par\end{centering}
\caption{ Summary of the optimised ROI sizes for the dedicated study of the LMC environment by varying the underlying IE model. The optimisation has been performed under the assumption of the LMC halo shapes according to the simulation in Ref.~\protect\cite{Donaldson+2022} using an Hernquist (MW) and NFW (LMC) profile as initial conditions. \label{tab:lmc_rois_iem}}
\end{table*}

\begin{table*}
\begin{center}
\begin{tabular}{lc}
\hline
IE model & optimal ROI size \\
\hline
\hline
Lorimer I & $\ell\in\left[-167^{\circ},167^{\circ}\right]$ and $b\in\left[-90^{\circ},-35^{\circ}\right]$  \\
Lorimer II & $\ell\in\left[33^{\circ},327^{\circ}\right]$ and $b\in\left[-90^{\circ},-30^{\circ}\right]$  \\
FGMA & $\ell\in\left[-165^{\circ},165^{\circ}\right]$ and $b\in\left[-90^{\circ},-35^{\circ}\right]$  \\
FGMB & $\ell\in\left[21^{\circ},339^{\circ}\right]$ and $b\in\left[-90^{\circ},-35^{\circ}\right]$  \\
FGMC & $\ell\in\left[25^{\circ},335^{\circ}\right]$ and $b\in\left[-90^{\circ},-35^{\circ}\right]$  \\
\hline
\end{tabular}
\caption{Summary of the optimised ROI sizes for the five IE models used in the analysis of the outer MW halo. These optimised ROIs are derived with respect to the spatial morphology of the MW-LMC system found in the Erkal19 including the dynamical response of both DM haloes. \label{tab:iem_rois}}
\end{center}
\end{table*}	

\section{Dark matter constraints for an alternative annihilation channel}
\label{app:tau_results}

\begin{figure*}
\centering
\includegraphics[width=0.49\textwidth]{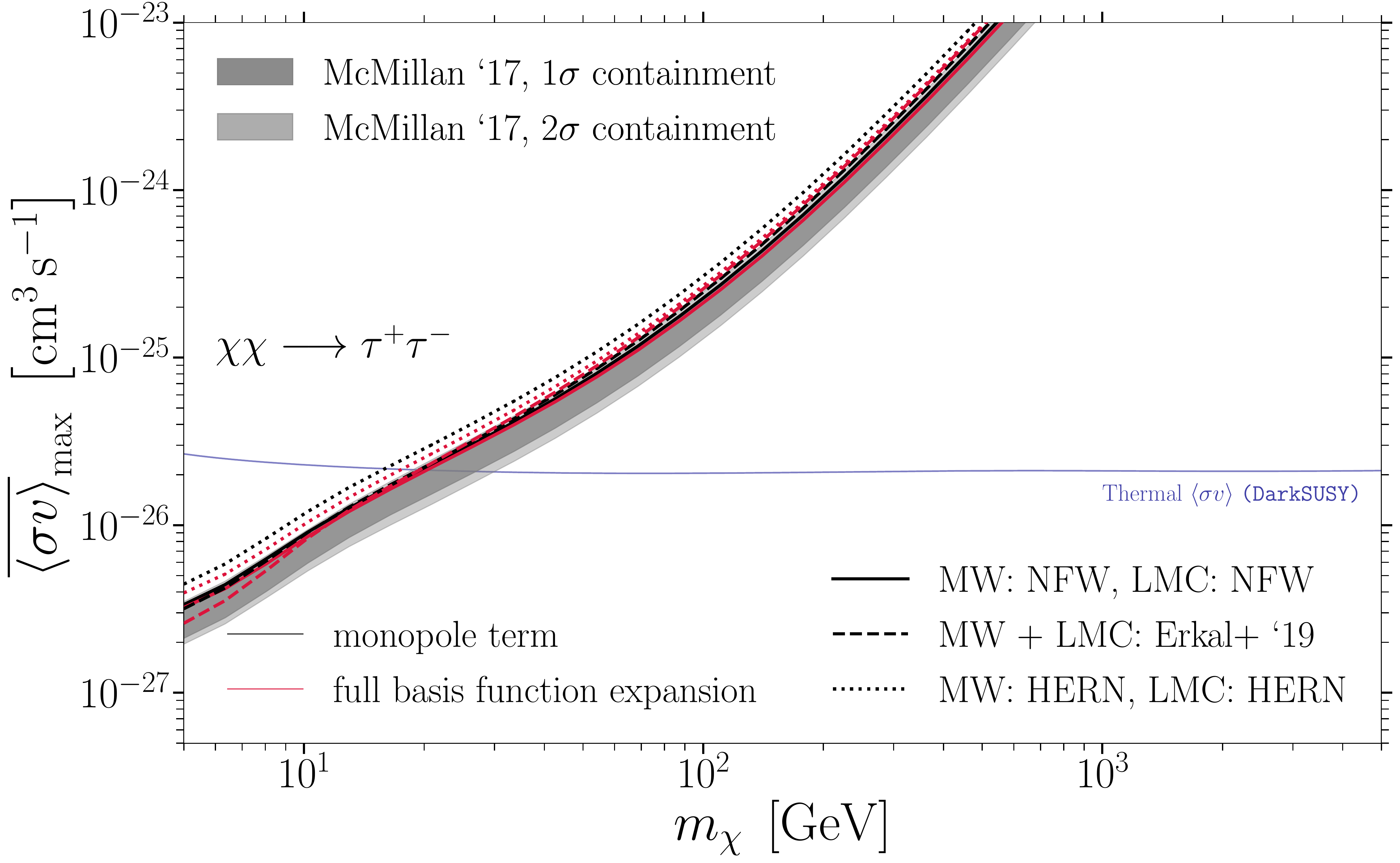}
\hfill
\includegraphics[width=0.49\textwidth]{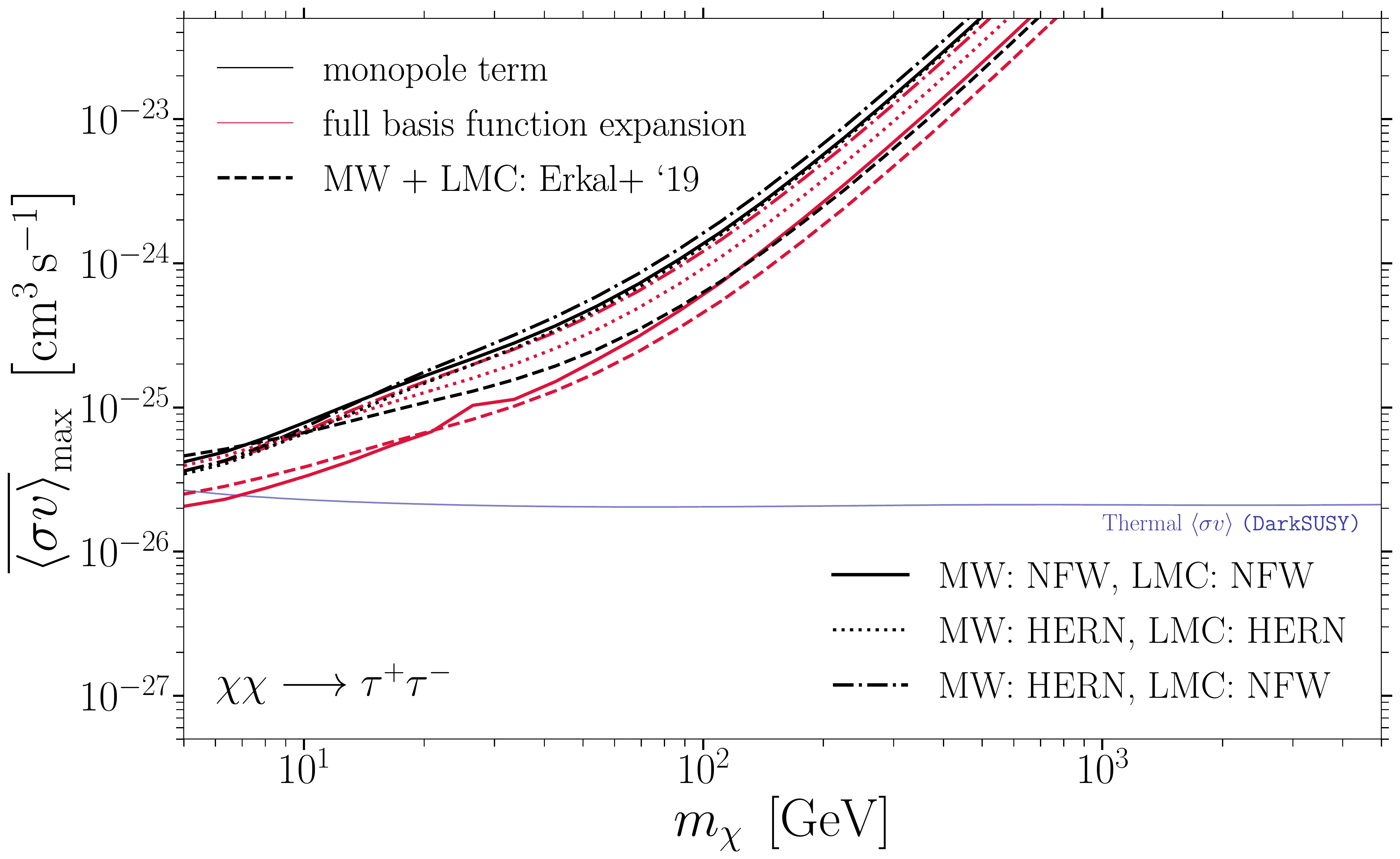}
\caption{(\emph{Left}:) Same as Fig.~\ref{fig:LMC_impact_on_MW} assuming DM pair-annihilation into $\tau^+\tau^-$ pairs. (\emph{Right}:) Same as the left panel of Fig.~\ref{fig:LMC_ROI_study} assuming DM pair-annihilation into $\tau^+\tau^-$ pairs.
\label{fig:results_tautau}}
\end{figure*}

In this appendix, we present our results for an alternative spectral DM model. Instead of the rather soft $b\bar{b}$-channel, we here assume 100\% annihilation into $\tau^+ \tau^-$ pairs that yield a hard annihilation spectrum. We repeat the analysis with respect to the optimised ROIs to re-derive the main results that have been shown in the main text in Fig.~\ref{fig:LMC_impact_on_MW} for the MW-LMC system and in Fig.~\ref{fig:LMC_ROI_study} for the LMC standalone. The corresponding constraints for the $\tau^+\tau^-$-channel are provided in Fig.~\ref{fig:results_tautau}.

As a general observation, the upper limits for DM particles with masses above a few tens of GeV are weaker compared to the $b\bar{b}$-channel. This result is reasonable given the fact that the $\tau^+ \tau^-$ annihilation spectrum is harder, which shifts the most constraining part of the spectrum to higher energies. At energies of a few GeV, \Fermi-LAT analyses tend to be statistics limited. As a consequence, their constraining power regarding hard spectra peaking above $\mathcal{O}(10)$ GeV is reduced.

\section{Dark matter constraints for an enlarged region of interest}
\label{app:newROI_limits}

In the main text of this work, we focused on determining an ROI in the southern hemisphere with optimal properties regarding the compatibility of the statistical expectations derived from our baseline fit and the true \Fermi-LAT data set. On one side, the dynamics of the MW-LMC system generate anisotropies that are located in this particular fraction of the entire sky as illustrated by Figs.~\ref{fig:jfactor_comparison} and \ref{fig:jfactor_comparison_LMC}. On the other side, such deformations -- mostly associated with the LMC halo -- are likewise present in the northern hemisphere. Their impact on indirect gamma-ray searches for DM in the outer MW halo remains unprobed.

\begin{figure*}
\centering
\includegraphics[width=0.49\textwidth]{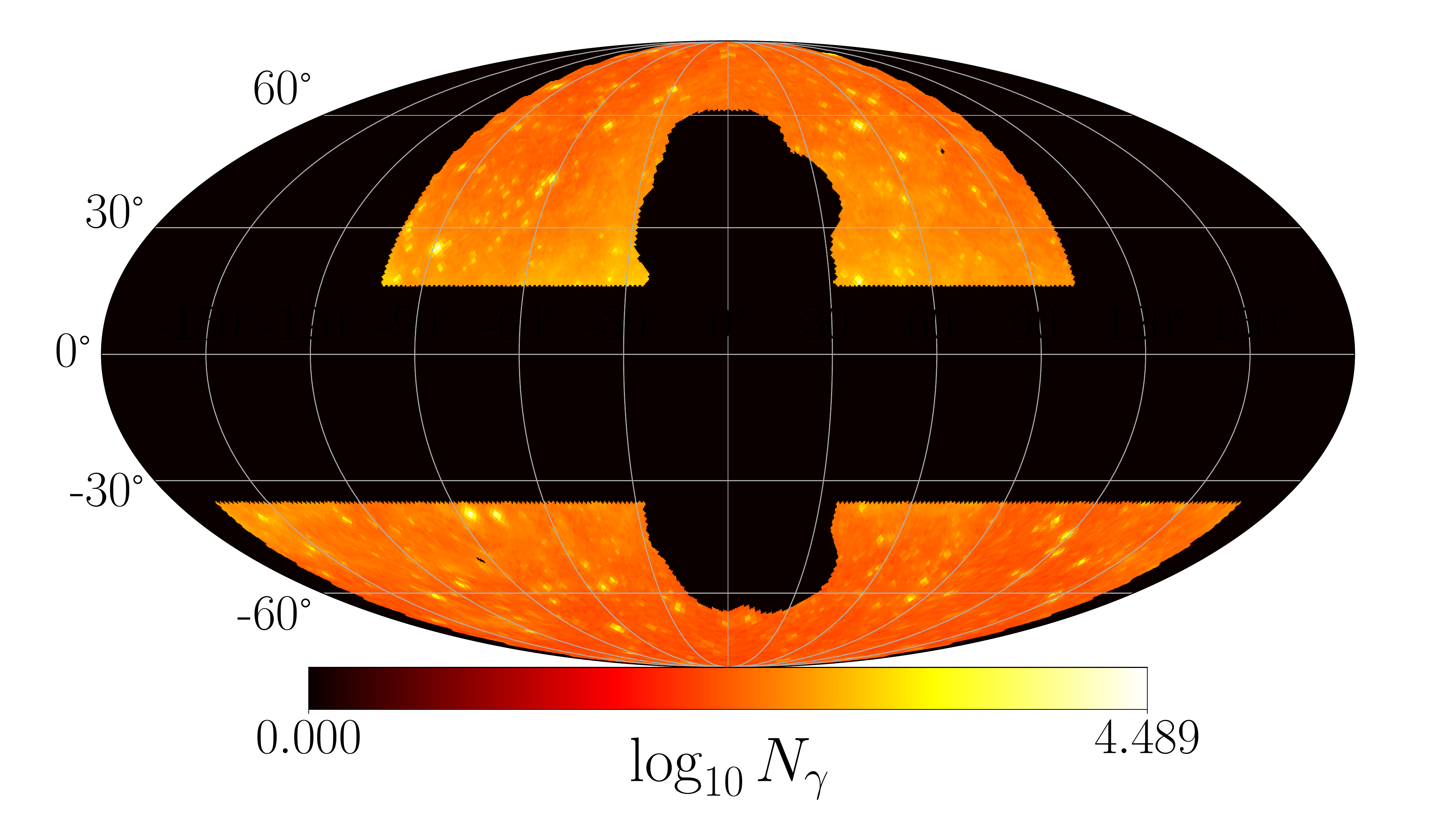}
\hfill
\includegraphics[width=0.49\textwidth]{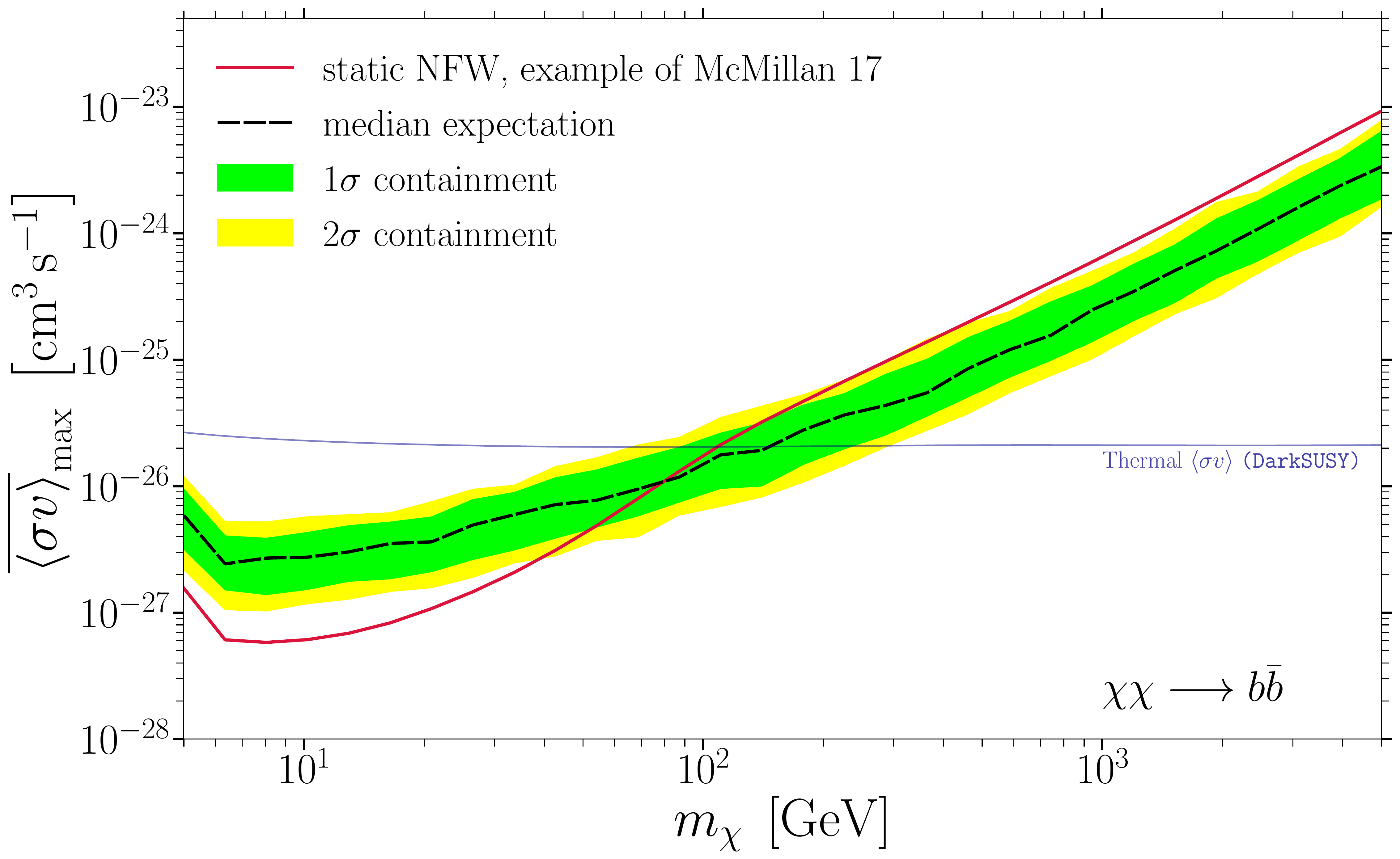}
\caption{(\emph{Left}:) Same as Fig.~\ref{fig:optimal_roi} including a fraction of the sky in the northern hemisphere defined by $-102^{\circ} < \ell < 102^{\circ}$ and $b > 16^{\circ}$ (excluding the FBs region). (\emph{Right}:) Same as the right panel of Fig.~\ref{fig:McMillan_example_ROI} regarding the ROI shown in the left panel.
\label{fig:McMillan_example_extended_ROI}}
\end{figure*}

To investigate the potential of an extended ROI that incorporates patches in both hemispheres of the gamma-ray sky, we fix the optimal southern ROI derived for the benchmark IE model Lorimer I and re-perform the optimisation routine in the northern sky. With respect to the example of a static MW halo referenced in Fig.~\ref{fig:McMillan_example_ROI}, we obtain the best match between expectation from mock data and performance on real \Fermi-LAT data for an ROI defined by $-102^{\circ} < \ell < 102^{\circ}$ and $b > 16^{\circ}$ (shown in the left panel of Fig.~\ref{fig:McMillan_example_extended_ROI}). The quality of this accordance is, however, not as good as for the southern hemisphere standalone. We quantify the suitability of the obtained extended ROI in the right pannel of Fig.~\ref{fig:McMillan_example_extended_ROI} by confronting the DM upper limits from true \Fermi-LAT data (red) with the corresponding constraints (black) and their scatter from mock data. The region in the parameter space where we can exclude thermal DM enlarges by a factor of about two compared to Fig.~\ref{fig:McMillan_example_ROI}. However, the derived bounds are not within the $2\sigma$ containment band of the upper limits derived from the baseline fit in a broad range of the probed DM parameter space. This reduces the credibility of this set of upper limits and it furthermore indicates that our baseline astrophysics model is not entirely describing reality in such an extended sky region. In order to exploit the full potential of searches towards the outer MW halo it is thus essential to improve the current state-of-the-art of astrophysical models at high latitudes; in particular, the gamma-ray emission of extended diffuse components as the FBs or loop-like structures and the residual IE.

As concerns the sensitivity of such indirect searches to deformations of the MW and LMC haloes, we find ROIs of similar extensions in the northern hemisphere for the Erkal19 simulation. The statistical robustness that these sky regions provide is comparable to the case shown in Fig.~\ref{fig:McMillan_example_extended_ROI}.  Translated to constraints on the DM annihilation cross section, we do not observe an enhanced difference between a signal morphology following the monopole term or the full BFE of the Erkal19 model. The effect of the MW-LMC dynamics remains at the level displayed in Fig.~\ref{fig:LMC_impact_on_MW}. To stress it again, the reliability of this statement is suffering from the lack of a good fit to reality. Improving the astrophysical background modelling at high latitudes may also improve the sensitivity to the dynamical response of the LMC passage through the MW.

\bsp	
\label{lastpage}
\end{document}